\documentclass[aps,pra,twocolumn,twoside,10pt]{revtex4-1}

\usepackage{graphicx,epstopdf,amsmath,amssymb,braket,color,epsfig,epstopdf,geometry,amsthm}
\usepackage[colorlinks=true, citecolor=red, urlcolor=blue ]{hyperref}
\usepackage[table]{xcolor}
\usepackage{times}

\newtheorem{proposition}{Proposition}
\newtheorem{corollary}{Corollary}[proposition]

\begin{document}

\title{Canonical Distillation of Entanglement
}
\author{
Tamoghna Das, Asutosh Kumar,  Amit Kumar Pal, Namrata Shukla, Aditi Sen(De), and Ujjwal Sen
}
\affiliation{Harish-Chandra Research Institute, Chhatnag Road, Jhunsi, Allahabad - 211019, India}
\affiliation{Homi Bhabha National Institute, Training School Complex, Anushaktinagar, Mumbai 400094, India}

\begin{abstract}
Distilling highly entangled quantum states from weaker ones is a process that is crucial for efficient and long-distance quantum communication, and has implications for several other quantum information protocols. We introduce the notion of distillation under limited resources, and specifically focus on the energy constraint. The corresponding protocol, which we call the canonical distillation of entanglement, naturally leads to the set of canonically distillable states. We show that for non-interacting Hamiltonians, almost no states are canonically distillable, while the situation can be drastically different for interacting ones. Several paradigmatic Hamiltonians are considered for bipartite as well as multipartite canonical distillability. The results have potential applications for practical quantum communication devices. 
\end{abstract}


\maketitle

\section{Introduction}
\label{intro}
Over the last twenty five years or so, entangled quantum states 
shared between distant parties have been proved to be essential for several quantum protocols \cite{ent_rmp,nature_briegel,ASDUS,diqc}.  
However, unavoidable destruction of quantum coherence due to noisy quantum channels diminishes the quality of the 
shared quantum state, thereby posing a challenge to the implementation of such protocols.  
Invention of distillation protocols \cite{distil,be,npptbe,prot_bennet} to 
purify highly entangled states from collection of states with relatively low entanglement has been proven crucial in order 
to overcome such difficulties in device independent quantum cryptography \cite{diqc,crypto}, quantum dense coding \cite{densecode}, 
and quantum teleportation \cite{teleport} - the three pillars 
of quantum communication. Entanglement distillation is also indispensable in quantum repeater
models \cite{repeater}, used to overcome the exponential scaling of the error probabilities with the length of the 
noisy quantum channel connecting distant parties sharing the quantum state. 
Existence of bound entangled (BE) states \cite{be} - entangled states from which no pure 
entangled state can be obtained using local operations and classical communications (LOCC) - further
highlights the importance of identifying distillable states. Entanglement distillation protocols have also been 
used in problems related to topological quantum memory \cite{qmem}.
Laboratory realization of single copy distillation has been performed and 
possible experimental proposal of multicopy distillation has been given \cite{expt}.

There is a close correspondence between entanglement and energy \cite{enten, ecost, eform, be}.
Moreover, consideration of statistical ensembles of quantum states of a system in terms of various constraints on its 
energy and number 
of particles is crucial in several areas of physics, including in quantum communication. 
An important example is the classical capacity of a noiseless quantum channel 
\cite{classchan,yuen,holevo,infen} for transmitting  
classical information using quantum states. The classical capacity is quantified by the von Neumann
entropy of the maximally mixed quantum state that can be sent through the noiseless quantum channel. 
The \textquotedblleft Holevo bound\textquotedblright \cite{classchan,yuen,holevo} dictates that at most $n$ bits of 
classical information can be transmitted using $n$ 
distinguishable qubits, thereby 
predicting an infinite capacity for infinite dimensional systems, such as the bosonic channels \cite{infen}. 
Since the energy required to 
achieve infinite capacity is also infinite, such non-physicality can be taken care of by calculating the capacity 
under appropriate energy constraints. Constraints on available energy can also be active in other quantum information protocols 
including infinite- as well as finite-dimensional systems and in particular may give rise to a novel understanding of the interplay 
between entanglement and energy. For example,  to implement 
ideas like quantum repeaters for long-distance quantum state distribution, 
an energy-constrained protocol for the  distillation of entanglement  may be necessary. 
Evidently, in that case, the energy of the states involved in the distillation process
must follow constraints according to the physical situation in hand, especially in the case of implementation of the 
protocol in the laboratory, where arbitrary amount of energy is not accessible. 
The logical choice of such constraints may include bounds on 
average energy, or maximum available energy of the quantum states.

In this paper, we consider the process of distillation of  highly entangled quantum states of shared systems from weakly entangled 
ones within the realm of limited resources. Specifically, we propose that a distillation protocol have to be carried 
out under an
energy constraint, and refer to it as \textquotedblleft canonical\textquotedblright distillation.
We prove that non-interacting Hamiltonians lead to  situations where canonically distillable states form a set of measure zero.
The 
situation, however, drastically changes with the inclusion of interaction terms. We consider several paradigmatic interacting 
Hamiltonians of spin-$\frac{1}{2}$ systems, viz. the transverse-field  $XY$ model \cite{xymodel,xybooks}, the longitudinal-field 
$XY$ model, and the 
$XXZ$ model in an applied field \cite{xxzmodel}, and the concept of canonical distillation is probed in each case. 
The interrelation between 
canonical distillability and the temperature in thermal states is also investigated. The findings are generic in the sense that they 
hold also in higher dimensions and for higher number of parties. The energy constraint in these cases is 
introduced by respectively considering a 
bilinear-biquadratic Hamiltonian \cite{bbh} 
of two spin-$1$ particles and a multisite transverse $XY$ model.  

The paper is organized as follows. In Sec. \ref{definition}, we define the canonical distillability of bipartite as well as 
multipartite quantum states. Sec. \ref{cd-biparty} contains the results on application of the canonical distillation protocol
in bipartite systems. The results are also demonstrated in the cases of well-known quantum spin models, where the canonical distillability of pure and mixed states with respect to these Hamiltonians are tested. In Sec. \ref{cd-multiparty}, we discuss the canonical distillability of multipartite states, 
focusing on three-qubit pure states belonging to the Greenberger-Horne-Zeilinger (GHZ) \cite{ghzstate,dvc} and the W \cite{dvc,zhgstate} classes. 
Sec. \ref{conclusion} contains the concluding remarks.

\begin{figure}
 \includegraphics[scale=0.4]{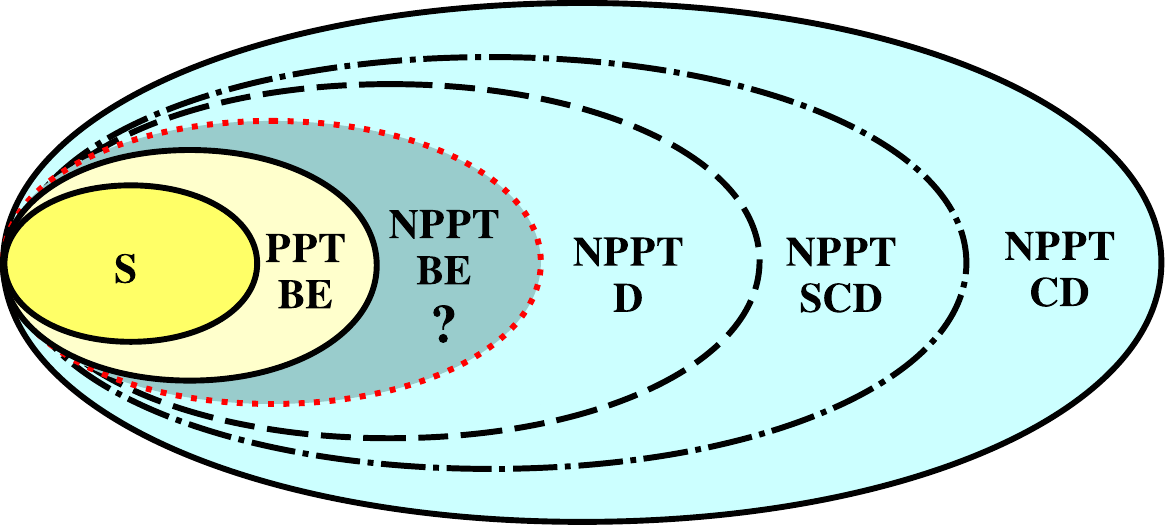}
 \caption{(Color online.) The structure of the state space in light of their distillability and canonical 
 distillability, assuming that 
 $\mathcal{S}_{CD}\subset\mathcal{S}_{SCD}$. Separable (S) states, 
 bound entangled (BE) states with positive partial transpose (PPT), and the conjectured set of NPPT BE
 \cite{be} states form the 
 set of undistillable states.The debatable existence of NPPT BE states is indicated 
 with a question mark. All the states that are outside the dotted line
 are distillable (D) in the usual, non-canonical sense. 
 The SCD states are outside the dashed line, while   
 the CD states are outside the dot-dashed boundary. 
 The boundary between the states that are SCD  and those that are not so (the dashed line) is 
 a non-convex one, while the boundary between the NPPT BE states, if existing, 
 and the NPPT distillable 
 states is non-convex under an assumption \cite{beconvx}. 
}
 \label{distil:phase}
\end{figure}

\section{Distillation under canonical energy constraint}
\label{definition}

We begin by providing a formal definition of 
\emph{canonical} distillation of entanglement for two-qubit systems in the asymptotic limit. Generalization to higher dimensions
and higher number of parties are considered later. 
In \textquotedblleft usual\textquotedblright\; entanglement distillation \cite{distil}, one 
intends to produce the largest number, \(m\), of copies of the maximally entangled Bell pair,
$|\psi^{-}\rangle= (|01\rangle-|10\rangle)/\sqrt{2}$,
starting from $n$ $(m\leq n)$ copies of an entangled two-qubit state, $\rho$, using only LOCC.   
Let us consider an LOCC on $n$ copies of the state $\rho$ that creates the state $\sigma$ which is close 
to $m$ copies of $|\psi^{-}\rangle$, or its local unitary equivalent, $|\tilde{\psi}^{-}\rangle=\mathcal{U}_{4}|\psi^{-}\rangle$, 
so that 
\begin{eqnarray}
 \underset{n\rightarrow\infty}{\mbox{lim}}\mbox{tr}(\sigma{\tilde{\sigma}}^{\otimes m})=1,
 \label{fidelity}
\end{eqnarray}
where $\tilde{\sigma}=|\tilde{\psi}^{-}\rangle\langle\tilde{\psi}^{-}|$. 
Here, $\mathcal{U}_{4}=U_{2}^{1}\otimes U_{2}^{2}$, with  
$U_{2}^{1}$  and $U_{2}^{2}$ being unitary operators on the qubit Hilbert space. 
The distillable entanglement is given by $E_{D}=\max\underset{n\rightarrow\infty}{\mbox{lim}}\frac{m}{n}$, where the maximum is 
over all LOCC protocols satisfying Eq. (\ref{fidelity}).

To introduce an appropriate energy constraint, 
suppose that the two-qubit quantum system in the state $\rho$ is described by the Hamiltonian $H$. 
Here, by ``system'', we mean the quantum system containing the set of $n$
resource states, in turn containing the set of $m$ output states of the distillation protocol. 
We assume that the system is in 
contact with a heat bath such that the average energies of the input and output states of the distillation protocol are equal.  
This average energy conservation leads to the constraint
$\mbox{tr}(\tilde{H}_n\rho^{\otimes n})=\mbox{tr}(\tilde{H}_m\sigma)
\approx\mbox{tr}(\tilde{H}_m\tilde{\sigma}^{\otimes m})$, with $\tilde{H}_j=\sum_{i=1}^{j}I^{\otimes i-1}\otimes H\otimes I^{\otimes j-i}$,
which implies
\begin{eqnarray}
\mbox{tr}(H\rho)=\frac{m}{n}\left(\mbox{tr}(H\tilde{\sigma})\right).
\label{energy_constraint}
\end{eqnarray}
Here we assume that $n$ is sufficiently large, so that $\mbox{tr}(\tilde{H}_m\sigma)$ can be approximated by 
$\mbox{tr}(\tilde{H}_m{\tilde{\sigma}}^{\otimes m})$. It can be shown, by virtue of Eq. (\ref{fidelity}), that the 
approximation is an equality for $n\rightarrow\infty$.

Note that we are assuming an insignificant contribution in average energy from the $n-m$ bipartite systems that are traced out, 
and any additional ancillary systems that are used and then discarded out during the LOCC protocol for the canonical distillation. 
Such energy dissipation channels can be incorporated into the definition, but leads to further intractability in the 
analysis. On the other hand, this assumption can be justified by noticing that the remnants 
after the application of a usual distillation protocol for creating singlet from pure two-qubit non-maximallly entangled
states \cite{prot_bennet}, $\alpha|00\rangle+\beta|11\rangle$, are of 
the form $|0\rangle^{\otimes n}_A|0\rangle^{\otimes n}_B$ and 
$|1\rangle^{\otimes n}_A|1\rangle^{\otimes n}_B$ with probabilities $|\alpha|^{2n}$ and $|\beta|^{2n}$, respectively, 
where $A$ and $B$ are the two parties. 
This contributes in average energy of the system by an 
amount $\delta_{E}$, where $\delta_{E}=n[|\alpha|^{2n}\langle0_A0_B|H|0_A0_B\rangle]+|\beta|^{2n}\langle1_A1_B|H|1_A1_B\rangle]$. 
Since $0\leq|\alpha|,|\beta|\leq1$, for $|\alpha|,|\beta|\neq0,1$,
$\delta_E\rightarrow 0$ as $n\rightarrow\infty$. 
We will discuss specific examples in the coming sections, 
where we consider several important and specific forms of the system Hamiltonian. 
In the limit $n\rightarrow\infty$, from Eq. (\ref{energy_constraint}), we have 
\begin{eqnarray} 
\mbox{tr}(H\rho)=\underset{n\rightarrow\infty}{\mbox{lim}}\frac{m}{n}{\mbox{tr}(H\tilde{\sigma})}.
\label{cd_actual}
\end{eqnarray}
The average energy constraint can also be replaced by a maximal available energy constraint, wherein we expect the broad qualitative
features, of the case where the average energy is considered, to be retained.
The canonically distillable entanglement, $E_{CD}$, is the maximum value of $\underset{n\rightarrow\infty}{\mbox{lim}}\frac{m}{n}$
that satisfies Eq. (\ref{cd_actual}) for some $\mathcal{U}_{4}$, and is consistent with Eq. (\ref{fidelity}). 
We call the states with a non-zero $E_{CD}$ to be canonically distillable (CD). 
One must note that for the two-qubit systems,
\begin{eqnarray}  
0\leq E_{CD}\leq E_{D}\leq 1.
\label{distillability:qubits}
\end{eqnarray}  

We would like to emphasize here that the canonical energy constraint, in the present problem, is imposed on the ensemble of quantum states over which the LOCC protocol is applied. The LOCC protocol can be modeled by an appropriate choice of following two Hamiltonians: (a) the Hamiltonian corresponding to the laboratory setting implementing the protocol in a real experiment and (b) the Hamiltonian modeling the interaction between the system and the laboratory environment in the same experiment. These choices do put novel constraints over the average energy of the source states. However, we assume that the system has already equilibriated with its laboratory environment, so that the average energy constraint applied to the source states takes into account the restrictions resulting from the application of the LOCC process. This assumption is in the same vein as that considered in the problem of ascertaining the capacity in the case of bosonic channels, as mentioned in Sec. \ref{intro}. The capacity of a bosonic channel without an energy constraint is infinite. While it is important to understand the bosonic channel capacity after an energy constraint is applied to the entire process of encoding, sending, and decoding of the channel states, a physically relevant bosonic channel capacity is obtained also by providing an energy constraint on the source states \cite{infen}. A similar example is provided by equilibrium statistical mechanics, where an average energy constraint on the system Hamiltonian provides useful information about the system’s thermodynamical quantities, that is independent of (i) the Hamiltonian of the bath and (ii) the system-bath Hamiltonian, for a large class of the two latter Hamiltonians (in (i) and (ii)) \cite{pathria}.

Just like $E_{D}$, determination of $E_{CD}$ under the canonical energy constraint is a difficult problem. However, 
significant insight on CD states can be obtained by considering a weaker version of the energy constraint, viz. 
\begin{eqnarray}
 \mbox{tr}(H\rho)=\mbox{tr}(H\tilde{\sigma}).
 \label{cdeq}
\end{eqnarray}
We refer to this as the weak canonical energy constraint (WCEC). 
Replacing Eq. (\ref{cd_actual}) by Eq. (\ref{cdeq}) leads us to the concept of \textquotedblleft special\textquotedblright 
CD (SCD) states. 

Note here that the relation between the set of CD and SCD states depends on the allowed values of
$\mbox{tr}(H\rho)$ and $\mbox{tr}(H\tilde{\sigma})$. While $\mbox{tr}(H\rho)$ is bounded within the range $[E_1,E_2]$, where $E_1$ and 
$E_2$ are respectively the minimum and maximum eigenvalues of $H$, $\mbox{tr}(H\tilde{\sigma})$ can have a different accessible range
$[\epsilon_1,\epsilon_2]$ 
due to the involvement of the free unitaries, $\mathcal{U}_4$, where 
\begin{eqnarray}
 \epsilon_{1}=\underset{\mathcal{U}_4}{\min}~\mbox{tr}(H\tilde{\sigma}),\nonumber \\
 \epsilon_{2}=\underset{\mathcal{U}_4}{\max}~\mbox{tr}(H\tilde{\sigma}).
 \label{res_range_actual}
\end{eqnarray}
One can consider two different situations. 
\textbf{(i)} The first situation is when $\epsilon_1=\epsilon_2$, where the range $[E_1,E_2]$ and 
$[\epsilon_1,\epsilon_2]$ has zero
overlap, thereby forbidding special canonical distillability of almost all quantum states. Similar result for canonical distillability 
follows from Eq. (\ref{cd_actual}). \textbf{(ii)} The second
situation arises when $\epsilon_1\neq \epsilon_2$, in which case special canonical distillability of a quantum state 
$\rho$ is guranteed by the value of $\mbox{tr}(H\rho)$ being in the common region of the ranges $[E_1,E_2]$ and 
$[\epsilon_1,\epsilon_2]$. Note here that the set of all SCD states is clearly a superset of the set of all CD states
by virtue of Eq. (\ref{cd_actual}).

The definition of canonical distillability can be extended to bipartite states of arbitrary dimensions, 
where the individual parties have dimensions $d>2$. In that case, the states $|\psi^{-}\rangle$ and 
$|\tilde{\psi}^{-}\rangle$
are replaced by $|\Phi\rangle$ and $|\tilde{\Phi}\rangle$, respectively, with 
\begin{eqnarray}
 |\Phi\rangle=\frac{1}{\sqrt{d}}\sum_{i=1}^{d}|i\rangle_{1}|i\rangle_{2}
 \label{qudit_state}
\end{eqnarray}
being a maximally entangled pure state in $\mathbb{C}^{d}\otimes\mathbb{C}^{d}$, 
and $|\tilde{\Phi}\rangle=\mathcal{U}_{d^{2}}|\Phi\rangle$. Here, 
$\mathcal{U}_{d^{2}}=U^{1}_{d}\otimes U^{2}_{d}$, 
with $U^{k}_{d}$ being an unitary operator on $\mathbb{C}^d$ for $k=1,2$, and $\{|i\rangle;i=1,\cdots,d\}$
forms the computational basis in $\mathbb{C}^d$. 
Since we are trying to create maximally entangled states in $\mathbb{C}^{d}\otimes\mathbb{C}^{d}$, the relations in 
Eq. (\ref{distillability:qubits}) are still valid. This is the case where the state $|\Phi\rangle$ is considered to have 
unit entanglement. If $|\Phi\rangle$ is considered to possess $\log_{2}d$ ebits of entanglement, all the expressions in 
Eq. (\ref{distillability:qubits}) need to be multiplied by $\log_{2}d$. 
A schematic diagram of the major divisions in the state 
space with respect to distillability is depicted in Fig. \ref{distil:phase}, where $\mathcal{S}_{CD}\subset\mathcal{S}_{SCD}$ is 
assumed, with $\mathcal{S}_{CD}$ and $\mathcal{S}_{SCD}$ representing the sets of CD and SCD states, respectively.

We conclude this section by pointing out that a multipartite extension of canonical distillation can be achieved by considering $n$ copies of an $N$-party state, $\rho_{N}^{\otimes n}$, from which $m$ copies of $|\Psi_{N}\rangle$, a certain pure state, or its local unitary equivalent 
$|\tilde{\Psi}_{N}\rangle$, can be created using LOCC under the canonical constraint. The choice of $|\Psi_N\rangle$ may not be unique in this case \cite{dvc}, and depends on the usefulness of that state in quantum information tasks. A demonstration of canonical distillability in multipartite scenario is presented in Sec. \ref{cd-multiparty}.

\section{Bipartite systems}
\label{cd-biparty}

The first result that we prove on canonical distillability of bipartite quantum states is for the case of a general non-interacting 
Hamiltonian $H_{l}$, defined on a system of two qudits, each having dimension $d$. The Hamiltonian is given by
\begin{eqnarray}
 H_{l}=\vec{\alpha}.\vec{\mathcal{S}}_{1}\otimes I_{2}+I_{1}\otimes\vec{\beta}.\vec{\mathcal{S}}_{2}.
 \label{loc-ham}
\end{eqnarray}   
Here, $\vec{\mathcal{S}}=\{\mathcal{S}^{x},\mathcal{S}^{y},\mathcal{S}^{z}\}$ with $\mathcal{S}^{i}$ ($i=x$, $y$, $z$) 
being the $d$-dimensional 
spin operators of a quantum spin-$j$ particle $(j=(d-1)/2)$, $\vec{\alpha}$ and $\vec{\beta}$ are unit vectors, $I$ denotes the 
identity operator in $\mathbb{C}^{d}$, and the subscripts $1$ and $2$ denote the two qudits.

\begin{proposition}
 For a system of two qudits described by a non-interacting Hamiltonian, 
almost no states are SCD.
\label{pr:localham}
\end{proposition}

\begin{proof}
The maximally entangled state, $|\Phi\rangle$ (Eq. (\ref{qudit_state})), and its local unitary equivalents have zero magnetizations 
in their single-party local density matrices. Therefore, 
\begin{eqnarray}
\langle\tilde{\Phi}|\mathcal{S}_{1}^{i}\otimes I_{2}|\tilde{\Phi}\rangle 
=\langle\tilde{\Phi}|I_{1}\otimes\mathcal{S}_{2}^{i}|\tilde{\Phi}\rangle=0
\label{loc-zero}
\end{eqnarray}
$\forall i=x$, $y$, $z$.  
Hence, for two-qudit Hamiltonians of the form (\ref{loc-ham}), the WCEC reduces to the form 
\begin{eqnarray}
 \mbox{tr}(H_{l}\rho)=0.
 \label{cd-hyperplane}
\end{eqnarray}
The probability that a state (pure or mixed) chosen randomly
from the entire state space to lie on this surface (Eq. (\ref{cd-hyperplane})) is vanishingly small. 
Hence, almost no 
two-qudit states are SCD if the system is described by a non-interacting Hamiltonian of the form $H_{l}$. 
\end{proof}

Due to Eq. (\ref{cd_actual}), Proposition \ref{pr:localham} immediately leads to the following corollary.

\begin{corollary}
 For a system of two qudits described by a non-interacting Hamiltonian, 
almost no states are CD.
\label{cor:localham}
\end{corollary}

\noindent\textbf{Note.} Proposition \ref{pr:localham} and Corollary \ref{cor:localham} 
are true also in the general case when the local parts of the Hamiltonian $H_{l}$
are expressed as linear combinations of generators of $SU(d)$.

To investigate whether introduction of interaction terms in the Hamiltonian has any effect on 
canonical distillability of the states of the system, we consider a Hamiltonian of the form $H=xH_{int}+yH_{l}$. 
Here, $H_{int}$ is the interacting part of $H$ whereas $H_{l}$ is the local part having a generic form as 
given in Eq. (\ref{loc-ham}), and $x$ and $y$ are appropriate system parameters.
Without any loss of generality, one can scale the system 
by the parameter $x$ so that the Hamiltonian of the system takes the form $H=H_{int}+gH_{l}$ with $g=y/x$.
Let us consider the minimalistic interacting Hamiltonian for the system of two qudits given by 
\begin{eqnarray}
 H_{int}=\vec{n}_{1}.\vec{\mathcal{S}}_{1}\otimes\vec{n}_{2}.\vec{\mathcal{S}}_{2},
 \label{smallinteraction}
\end{eqnarray}
where $\vec{n}_{1}$ and $\vec{n}_{2}$ are unit vectors. We refer to $\frac{1}{g}$ as the 
``participation ratio'' of the interaction part, $H_{int}$, 
in the Hamiltonian $H$, with respect to the local part, $H_{l}$.

\begin{proposition}
 Introduction of the minimalistic interacting part in the Hamiltonian, with arbitrarily small participation 
ratio with respect to the local part, results in a
non-zero probability of a randomly chosen state to be SCD.
\label{pr:int_term}
\end{proposition}

\begin{proof}
For a system of two qudits described by the Hamiltonian $H=H_{int}+gH_{l}$, where $H_{int}$ and 
$H_{l}$ are given by Eqs. (\ref{smallinteraction}) and (\ref{loc-ham}), respectively, the WCEC reduces to 
\begin{eqnarray}
 \mbox{tr}(H\rho)=\langle\tilde{\Phi}|H_{int}|\tilde{\Phi}\rangle, 
\end{eqnarray}
where we have used Eq. (\ref{loc-zero}). 
Suppose that the limits of variation of the quantity, $\mbox{tr}(H\rho)$, of the WCEC for a specific value of $g$ 
are $E_{1}^{g}$ and $E_{2}^{g}$, i.e., $\mbox{tr}(H\rho)\in[E_{1}^{g},E_{2}^{g}]$. 
Note that $E_{1}^{g}$ and $E_{2}^{g}$ respectively are the minimum 
and the maximum eigenvalues of $H$ for a specific value of $g$, while let $E_{1}^{int}$ and $E_{2}^{int}$ be the same of $H_{int}$. 
The accessible range of the right hand side of the WCEC is given by 
$\langle\tilde{\Phi}|H_{int}|\tilde{\Phi}\rangle\in[\epsilon_{1},\epsilon_{2}]$, where, according to Eq. (\ref{res_range_actual}),
\begin{eqnarray}
\epsilon_{1}&=&\underset{\{U_{d}^{k}\}}{\min}\langle\tilde{\Phi}|H_{int}|\tilde{\Phi}\rangle,\nonumber \\ 
\epsilon_{2}&=&\underset{\{U_{d}^{k}\}}{\max}\langle\tilde{\Phi}|H_{int}|\tilde{\Phi}\rangle,
\label{int_range_actual}
\end{eqnarray}
with $E_{1}^{int}\leq\epsilon_{1}\leq\epsilon_{2}\leq E_{2}^{int}$. 
Evidently, the probability of a randomly chosen two-qudit state to 
be SCD is non-zero iff \textit{(i)} $\epsilon_{1}\neq\epsilon_{2}$, and \textit{(ii)} 
$[\epsilon_{1},\epsilon_{2}]$ and $[E_{1}^{g},E_{2}^{g}]$ have a non-zero overlap. 

Considering the form (\ref{smallinteraction}) of $H_{int}$, $\vec{n}_{i}.\vec{\mathcal{S}}_{i}$ is the component of the 
spin-$j$ operator, $\vec{\mathcal{S}}_{i}$, along the direction $\vec{n}_{i}$, and let its eigenvectors be $\{|m^{j}\rangle_{i,\vec{n}_{i}}\}$. 
Let $\{|m^{j}\rangle_{i}\}$ be the eigenvectors of $\mathcal{S}_{i}^{z}$ with the eigenvalues $\{m^{j}_{i}\}$. In this basis, the 
state given in Eq. (\ref{qudit_state}) can be replaced by 
\begin{eqnarray}
 |\Phi\rangle=\frac{1}{\sqrt{2j+1}}\sum_{m^{j}=-j}^{j}|m^{j}\rangle_{1}|m^{j}\rangle_{2}.
 \label{qudit_state_2}
\end{eqnarray}
A convenient choice of unitary operators 
is where $U_{1}^{d}|m^{j}\rangle_{1}=|m^{j}\rangle_{1,\vec{n}_{1}}$ and 
$U_{2}^{d}|m^{j}\rangle_{1}=|m^{j}\rangle_{2,\vec{n}_{2}}$, which results in 
\begin{eqnarray}
 \langle\tilde{\Phi}|H_{int}|\tilde{\Phi}\rangle=\frac{1}{2j+1}\sum_{m^{j}=-j}^{j}(m^{j})^{2}=\frac{1}{3}j(j+1).
\end{eqnarray}
Similarly, for a different choice of unitary operators, viz. $V_{1}^{d}|m^{j}\rangle_{1}=|m^{j}\rangle_{1,\vec{n}_{1}}$
and $V_{2}^{d}|m^{j}\rangle_{1}=|-m^{j}\rangle_{2,\vec{n}_{2}}$, one can obtain 
$\langle\tilde{\Phi}|H_{int}|\tilde{\Phi}\rangle=-\frac{1}{3}j(j+1)$. From Eq. (\ref{int_range_actual}), 
$\epsilon_{1}\leq -\frac{1}{3}j(j+1)$ and $\epsilon_{2}\geq\frac{1}{3}j(j+1)$, thereby proving $\epsilon_{1}\neq\epsilon_{2}$.

To prove condition \textit{(ii)}, we point out that the Hamiltonian $H$ is traceless, which implies $E_{1}^{g}<0$ and $E_{2}^{g}>0$
for a specific value of $g$. From the above discussion, it is proved that $\epsilon_{1}<0$ and $\epsilon_{2}>0$, which is possible 
only when the ranges $[\epsilon_{1},\epsilon_{2}]$ and $[E_{1}^{g},E_{2}^{g}]$ have a finite overlap. 
Hence, the proof. 
\end{proof}

\noindent Corollary \ref{cor:int_term_qubit} follows directly from Proposition \ref{pr:int_term}, with an additional feature.

\begin{corollary}
 For a system of two qubits described by a Hamiltonian of the form $H=H_{int}+gH_{l}$, 
a non-zero probability for a randomly chosen two-qubit state to be SCD is guaranteed by a finite overlap of the ranges $[E_{1}^{g},E_{2}^{g}]$ and 
$[E_{1}^{int},E_{2}^{int}]$ $=$ $[\epsilon_{1},\epsilon_{2}]$.
\label{cor:int_term_qubit}
\end{corollary}

\begin{proof}
For a system of two qudits, considering the form of $H_{int}$ given in Eq. (\ref{smallinteraction}), 
$E_{1}^{int}=-E_{2}^{int}=-j^{2}$. For a two-qubit system ($j=\frac{1}{2}$), $\epsilon_{1}=-\epsilon_{2}=\frac{1}{4}$, implying 
$\epsilon_{1}=E_{1}^{int}$ and $\epsilon_{2}=E_{2}^{int}$. Therefore, varying the unitary operators, one can exhaust the full range of the right hand
side of WCEC. Having $[E_{1}^{int},E_{2}^{int}]$ $=$ $[\epsilon_{1},\epsilon_{2}]$ is the additional feature in 
Corollary \ref{cor:int_term_qubit} with respect to 
Proposition \ref{pr:int_term}. 
\end{proof}

\noindent Note that for Hamiltonians of the form $H=H_{int}+gH_l$, the value of $\delta_E$ depends on the Hamiltonian parameter 
$g$. For example, if one considers $H_{int}=\vec{S}_1\otimes\vec{S}_{2}$ and $H_l=S_1^z\otimes I+I\otimes S_2^{z}$, then 
$\delta_E=n[|\alpha|^{2n}(g+\frac{1}{4})+|\beta|^{2n}(g+\frac{1}{4})]$ tends to zero for large $n$ when $0<|\alpha|,|\beta|<1$.

Next, we wish to estimate the probability that a given quantum state, $\rho$, is SCD with 
respect to the Hamiltonian $H$ for a specific value of $g$. If the states $\rho$
are uniformly distributed in the energy range $[E_{1}^{g},E_{2}^{g}]$, the required probability would just be the ratio 
of the lengths of the two energy ranges, $[E_{1}^{g},E_{2}^{g}]$ and $[\epsilon_{1},\epsilon_{2}]$. 
This, however, is not the case, and the states, $\rho$, for any given rank, $r$, are typically distributed on the energy range 
$[E_{1}^{g},E_{2}^{g}]$ with a bell-shape. 

Let $P(E)dE$ denotes the probability that an arbitrary two-qudit state $\rho$ of rank $r$ has average energy between 
$E$ and $E+dE$. To keep the notations uncluttered, the symbol $r$ is not included in the probability density function $P(E)$. 
Let $P(\mbox{dist}|E)$ be the probability density that the state $\rho$ (of rank $r$) is distillable (in the usual, non-canonical
sense) given that its average energy is $E$. Therefore, the probability, $p$, that a given state, $\rho$, of rank $r$, is SCD 
with respect to the Hamiltonian $H$ is given by 
\begin{eqnarray}
 p=\int_{\epsilon_{1}}^{\epsilon_{2}}P(\mbox{dist}|E)P(E)dE.
 \label{p:general}
\end{eqnarray}

There does not, as yet, exist an efficient method to estimate the quantity $P(\mbox{dist}|E)$ in arbitrary dimensions. However, 
in $\mathbb{C}^{2}\otimes\mathbb{C}^{d}$ systems, distillability is equivalent to being non-positive under partial 
transposition (NPPT) \cite{ppt-ph}. For such systems, we can perform numerical simulations to estimate this quantity. Interestingly, in all 
the systems that we have considered, numerical evidence indicates that $P(\mbox{dist}|E)$ is independent of $E$. We will refer to 
this as the \textquotedblleft assumption of independence\textquotedblright (AI), and denote $P(\mbox{dist}|E)$ by $\eta$ when the 
assumption is valid. We refer to $\eta$ as the distillability factor (DF). 
We have, therefore, the following proposition under the assumption of 
independence.

\begin{proposition}
 The probability of an arbitrary entangled state, pure or mixed, of a 
two-qudit system defined by the Hamiltonian $H=H_{int}+gH_{l}$ for a specific value of $g$ to be SCD is given by
\begin{eqnarray}
 p=\eta\int_{\epsilon_{1}}^{\epsilon_{2}}P(E)dE. 
\end{eqnarray}
\label{pr:probscd}
\end{proposition}

\noindent\textbf{Note.} None of the results presented in the subsequent discussions uses the assumption of independence for the numerical calculations.  However, we expect that the formula in Proposition \ref{pr:probscd} will be useful in cases  where numerical simulation is more challenging. 

\subsection{Canonical distillation in quantum spin models}
\label{sec:qsm}

Now we apply the above formulation of canonical distillation of entanglement in the case of well-known quanum spin Hamiltonians, and discuss a number of interesting features of the special canonical distillability in these models. We start with $\mathbb{C}^2\otimes\mathbb{C}^2$ systems, where the spin operators are the Pauli matrices, $\{\sigma_{i}^{x}, \sigma_{i}^{y},\sigma_{i}^{z}\}$, acting on the qubit $i$. Following Proposition \ref{pr:localham}, there is a vanishing probability that a randomly chosen two-qubit state of a system described by a non-interacting two-qubit Hamiltonian of the form $H_{l}=\vec{\alpha}.\vec{\sigma}_{1}\otimes I_{2}+I_{1}\otimes\vec{\beta}.\vec{\sigma}_{2}$ with $\vec{\sigma_{i}}=\{\sigma_{i}^{x}, \sigma_{i}^{y},\sigma_{i}^{z}\}$ is special canonically distillable (SCD). 

We now consider the  two-qubit $XY$ Hamiltonian in a transverse-field \cite{xymodel,xybooks}, given by 
\begin{eqnarray}
H_{XY}=J(\gamma_{+}\sigma^{x}_{1}\sigma^{x}_{2}+\gamma_{-}\sigma^{y}_{1}\sigma^{y}_{2})
 +h\sum_{i=1}^2\sigma_{i}^{z},
 \label{xy_ham}
 \end{eqnarray} 
 with $\gamma_{\pm}=(1\pm\gamma)/2$
 representing the anisotropy and 
 $h$ being the field strength, can be expressed in the form $H_{XY}^{int}+gH_{XY}^l$, with $g=h/J$ (see \cite{supple}).   The existence of at least one unitary operator $\mathcal{U}_{4}$ for every value of $\langle\tilde{\psi}^{-}|H_{XY}^{int}|\tilde{\psi}^{-}\rangle$ in the allowed range $[-\epsilon,\epsilon]$, where $\pm\epsilon$ are the maximum and minimum eigenvalues of $H_{XY}^{int}$, results in a significant fraction of two-qubit SCD states for $J\neq0$ (Proposition \ref{pr:int_term}), while for $J=0$, almost all the states are not SCD (Proposition \ref{pr:localham}). Extensive numerical simulations suggest that AI holds within numerical accuracy, allowing one to calculate  $p$  with respect to $H_{XY}$ following Proposition \ref{pr:probscd}. The behaviour of $p$ at the extremums of the model is presented by the following Proposition (\cite{supple}).

\noindent\textbf{Proposition IV.} In a two-qubit system described by the transverse \(XY\) Hamiltonian, 
all entangled states are SCD, provided either \(h \rightarrow 0\) or \(\gamma \rightarrow \infty\).

The probability distributions, $P(E)$, of average energy of states over the space of all two-qubit pure states for different values of h/J are typically bell-shaped, and are determined by Haar-uniformly generating a sample of $10^8$ such states.  In corroboration with Proposition \ref{pr:localham}, for two-qubit pure states for which $\eta=1$, $p$ decreases as $h/J$ increases (see \cite{supple} for a depiction), and asymptotically vanishes as $h\rightarrow\infty$. For two-qubit mixed states, $\eta$ is a decreasing function of the rank, $r$, of the state, with $\eta(r=2)=1$, $\eta(r=3)=0.928$, and $\eta(r=4)=0.756$, estimated via numerical simulation by  generating $10^{8}$ states (for each rank) Haar uniformly over the space of quantum states of the corresponding rank. Similar to the pure states, $P(E)$ for the two-qubit entangled mixed states with different ranks are also bell-shaped with $p\rightarrow0$ for $h\rightarrow\infty$, as depicted in \cite{supple}. These results for the pure and mixed two-qubit states qualitatively hold also for a two-qubit $XXZ$ Hamiltonian in an external field, given by \cite{xxzmodel}
\begin{eqnarray}
 H_{XXZ}=\frac{J}{2}(\sigma_{1}^{x}\sigma_{2}^{x}+\sigma_{1}^{y}\sigma_{2}^{y}+\Delta\sigma_{1}^{z}\sigma_{2}^{z})
  +h\sum_{i=1}^{2}\sigma_{i}^{z},
 \label{xxz-ham_main}
\end{eqnarray} 
or a higher-dimensional system such as a bilinear-biquadratic spin Hamiltonian \cite{bbh}, defined on two qutrits and expressed as 
\begin{eqnarray}
 H_{3,3}=J\left[\cos\theta\vec{S}_{1}.\vec{S}_{2}+\sin\theta\left(\vec{S}_{1}.\vec{S}_{2}\right)^{2}\right]
 +h\sum_{i=1}^{2}S_{i}^{z},
\end{eqnarray}
where $\vec{S}_{i}=\left\{S_{i}^{x},S_{i}^{y},S_{i}^{z}\right\}$, $i=1,2$, are the spin operators on the qutrit $i$, with 
\begin{eqnarray}
 S_{i}^{x}&=&\frac{1}{\sqrt{2}}
 \left( 
 \begin{array}{ccc}
  0 & 1 & 0 \\
  1 & 0 & 1 \\
  0 & 1 & 0
 \end{array}
 \right),
  S_{i}^{y}=\frac{1}{\sqrt{2}i}
 \left( 
 \begin{array}{ccc}
  0 & 1 & 0 \\
  -1 & 0 & 1 \\
  0 & -1 & 0
 \end{array}
 \right), \nonumber \\
  S_{i}^{z}&=&\mbox{diag}\{1,0,-1\}.
\end{eqnarray}
Here, $\cos\theta$ and $\sin\theta$ are the relative strengths of the bilinear and biquadratic interactions, respectively. 
Note that the value of $p$ can be enhanced by changing the direction of the external  field, as in the case of the two-qubit $XY$ model in a longitudinal field, whose Hamiltonian is given by Eq. (\ref{xy_ham}), with $\sigma^z$ replaced by $\sigma^x$ in the local part.

Interestingly, there exists a critical \emph{SCD temperature} for every value $h/J$ in the XY as well as XXZ models, 
above which a two-qubit mixed thermal state  satisfies the WCEC, whereas below the critical value, it does not. 
Although an increase in the entanglement facilitates canonical distillability at zero temperature, after thermal mixing, 
there is a trade-off between temperature and entanglement which aids in canonical distillability at higher temperatures 
where entanglement is typically low. This is indicated by the existence of thermal states at high temperature having negligible or zero 
concurrence \cite{eform} but satisfying Eq. (\ref{cdeq}), and states having high thermal concurrence and yet violating WCEC \cite{supple}. 

\begin{figure} 
 \includegraphics[scale=0.35]{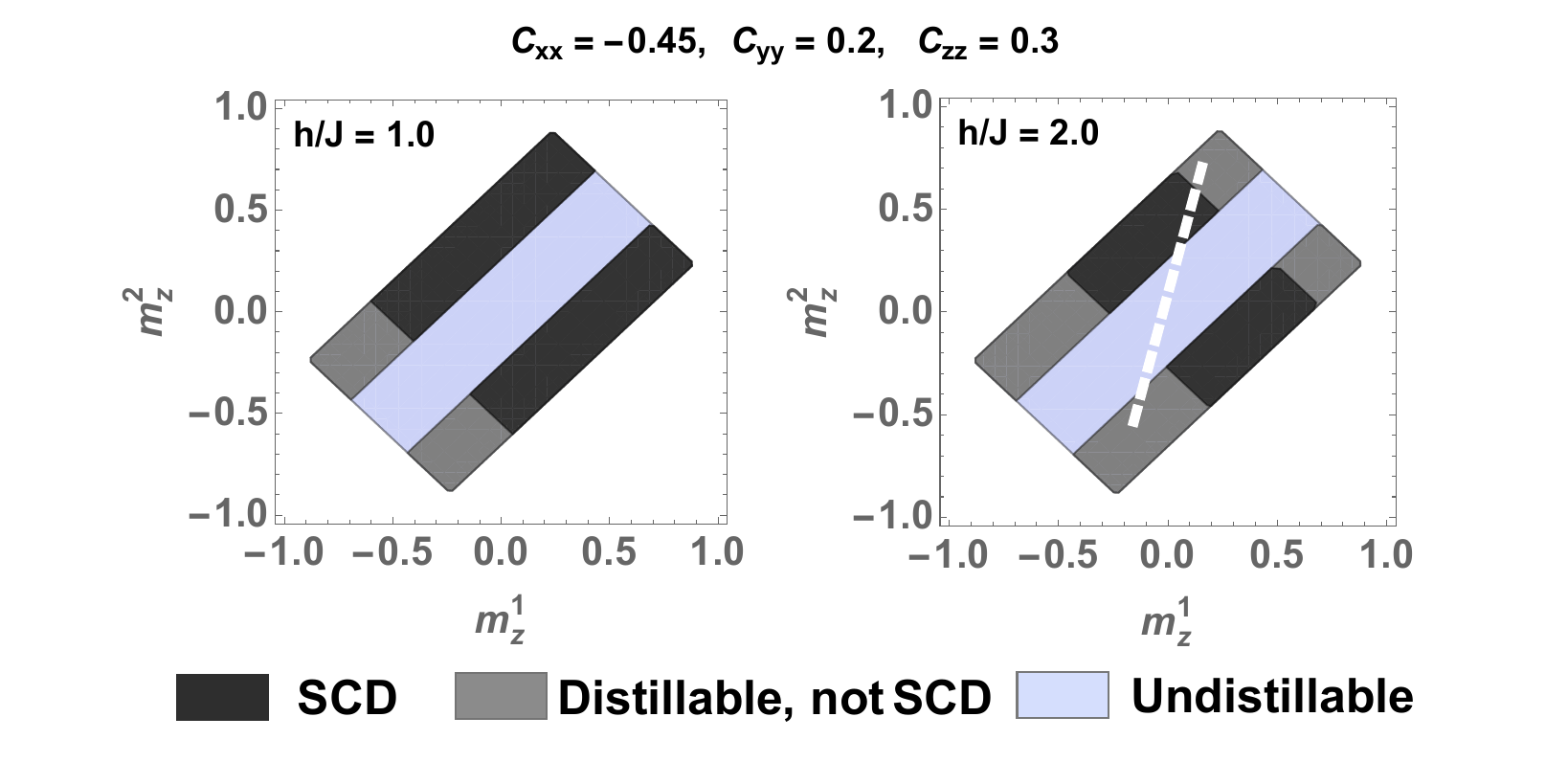}
 \caption{Canonical distillability of two-qubit states with local magnetizations. 
 The projections of SCD region on the $(m_{z}^{1},m_{z}^{2})$ plane of the 
 parameter space of the state \(\rho_{m}\), where the system is governed by the two-qubit \(XY\) Hamiltonian, is shown 
 by the black regions. The left figure is for $h/J=1.0$, while the right one is for $h/J=2.0$. 
 Note also that an SCD state can be obtained by mixing two quantum state that are distillable, but not SCD (the dark gray
 regions), as indicated by the dotted line, implying the non-convexity of the SCD states. The light gray regions indicate the 
 quantum states $\rho_{m}$ that are undistillable, whereas the white regions are for non-physical states. All axes are dimensionless.}
 \label{bellvolume_main}
 \end{figure}

The effect of magnetization of a two-qubit mixed state on canonical distillability becomes prominent in the case of  a two-qubit mixed state, $\rho_m$, constituted by only the diagonal elements of the correlation matrix, $c_{\alpha\alpha}=\mbox{tr}(\sigma_{1}^{\alpha}\otimes\sigma_{2}^{\alpha}\rho_m)$, $\alpha=x,y,z$, and the $z$ magnetization $m_{z}^{i}=\mbox{tr}(\sigma^{z}_{i}\rho_{i})$ \cite{supple},  with $|c_{\alpha\alpha}|,|m_z|\leq1$, $\rho_{i}$ being the local density matrix of the qubit $i$ ($i=1,2$). Canonical distillability of $\rho_m$ w.r.t. the two-qubit transverse-field XY model depends explicitly on the values  of $c_{\alpha\alpha}$ and $m_z$ (Fig. \ref{bellvolume_main}). In support of Proposition \ref{pr:localham}, a shrink in the SCD region with increasing $h/J$ indicates a decrease in the value of $p$ for a mixed state of the form $\rho_{m}$. Another illustration of this phenomena on the $(c_{xx},m_{z}^{2})$ plane can be found in \cite{supple}. Note that under the same Hamiltonian $H_{XY}$, Bell-diagonal states ($m_{z}^{i}=0$, $i=1,2$) are always SCD \cite{supple}.

\section{Multipartite systems}
\label{cd-multiparty}

As mentioned in Sec. \ref{cd-biparty}, canonical distillability of a multipartite system constituted of $N$ parties and described by a Hamiltonian $H$ 
can be investigated using the same methodology as that used in the case of bipartite systems. 
For the purpose of demonstration, we restrict ourselves to three-qubit states 
belonging to the well-known  GHZ and W classes \cite{ghzstate,zhgstate,dvc}, which are mutually disjoint sets that collectively 
exhaust the entire set of three-qubit pure states. Starting with three-qubit states chosen from the GHZ class as resource states, 
we consider the distillation of the multiparty entangled three-qubit GHZ state, 
$|\psi\rangle_{GHZ}=(|000\rangle+|111\rangle)/\sqrt{2}$, using the WCEC. Similarly, while considering the W 
class of states as the resource states, the target state is the three-qubit W state, 
$|\psi\rangle_{W}=(|001\rangle+|010\rangle+|100\rangle)/\sqrt{3}$. 
Let the multiqubit system be described by the $XY$ Hamiltonian in an external transverse field, given by 
\begin{eqnarray}
 H_{XY}^{N}&=&J\sum_{i=1}^{N}\left[\gamma_+\sigma_{i}^{x}\sigma_{i+1}^{x}
 +\gamma_-\sigma_{i}^{y}\sigma_{i+1}^{y}\right]
 +h\sum_{i=1}^{N}\sigma_{i}^{z}.
 \label{xy_ham_n_main}
\end{eqnarray}
Similar to its bipartite counterpart, in the case of $N=3$ also, the probability $p$ is determined as a decreasing function of $h/J$, for both GHZ and W class of states. Interestingly, for the W class states, $p$ is found to be unity up to $h/J=1$ and decreasing when $h/J>1$, whereas 
it decreases for the entire range of $h/J$ in the case of GHZ states (see \cite{supple} for depiction). 
This is therefore an occasion where it is advantageous to create W-class states than GHZ-class states (cf. \cite{W-vs-GHZ}).

\section{Conclusion} 
\label{conclusion}

Distillation of entanglement from shared quantum states is a useful technique for several quantum information protocols. It is 
important, for both fundamental and practical reasons, to consider this protocol in a scenario of limited resources. We 
have considered distillation of entanglement in bipartite and multipartite quantum states in the situation where there is a limited 
amount of energy that is accessible for the process to be accomplished. In particular, we consider constraints on average energy: 
a canonical energy constraint and a weak canonical energy constraint, that naturally lead to the concepts of canonical 
distillability and special canonical distillability. 
We have shown that for a bipartite system described by a non-interacting Hamiltonian, almost
no states are special canonically distillable. 
Significant understanding about the set of special canonically distillable states can be obtained by looking at the 
probability distributions of the 
average energies of the shared states. The concept has been applied to  a number of bipartite and multipartite systems described
by well-known spin Hamiltonians, namely, the spin-$\frac{1}{2}$ $XY$ model in transverse and longitudinal fields, the 
spin-$\frac{1}{2}$ $XXZ$ model in an external field, and the bilinear-biquadratic Hamiltonian of a spin-$1$ system in the 
presence of an external field.  
We find that the probability that a randomly chosen state is special canonically distillable can be manipulated 
by altering the direction of 
the external field.  The canonical distillability of a number of mixed states such as the thermal state, the Bell-diagonal states, 
and mixed bipartite states with fixed magnetizations are also investigated. It has been shown that for a fixed external 
field value, the thermal state can be special canonical distillable only above a critical temperature, which we have  
called the special canonically distillable temperature. The concept of canonical 
distillability of three-qubit GHZ and W class states have also been introduced. The results are expected to be of importance 
in realization of quantum communication channels.

\pagebreak
\widetext
\begin{center}
\textbf{\Large Supplementary Materials\\ \large Canonical distillation of entanglement} \\
\normalsize Tamoghna Das\(^{1,2}\), Asutosh Kumar\(^{1,2}\), Amit Kumar Pal\(^{1,2}\), Namrata Shukla\(^{1,2}\), Aditi Sen(De)\(^{1,2}\), and Ujjwal Sen\(^{1,2}\)\\
\small \(^1\)Harish-Chandra Research Institute, Chhatnag Road, Jhunsi, Allahabad 211019, India \\
\(^2\)Homi Bhabha National Institute, Training School Complex, Anushaktinagar, Mumbai 400094, India
\end{center}

\setcounter{equation}{0}
\setcounter{figure}{0}
\setcounter{table}{0}
\setcounter{section}{0}
\setcounter{page}{1}
\makeatletter
\renewcommand{\theequation}{SEQ\arabic{equation}}
\renewcommand{\thefigure}{SF\arabic{figure}}
\renewcommand{\thetable}{ST\arabic{table}}
\renewcommand{\bibnumfmt}[1]{[SR#1]}
\renewcommand{\citenumfont}[1]{SR#1}
\renewcommand{\thesection}{SSEC\arabic{section}}
\hypersetup{pageanchor=false}

\section{Two-Qubit Systems}
\label{cd:2qubit}

In the case of $d=2$, the spin operators are the Pauli matrices, $\{\sigma_{i}^{x}, \sigma_{i}^{y},\sigma_{i}^{z}\}$, acting 
on the qubit $i$. Following Proposition \ref{pr:localham}, there is a vanishing probability that
a randomly chosen two-qubit state of a system described by a non-interacting two-qubit Hamiltonian of the form 
$H_{l}=\vec{\alpha}.\vec{\sigma}_{1}\otimes I_{2}+I_{1}\otimes\vec{\beta}.\vec{\sigma}_{2}$ with $\vec{\sigma_{i}}=
\{\sigma_{i}^{x}, \sigma_{i}^{y},\sigma_{i}^{z}\}$ is special canonically distillable (SCD). 
Now we consider some examples of well-known two-qubit interacting Hamiltonians and investigate the canonical distillability of 
entangled states of two-qubit systems described by these Hamiltonians. 
We consider three well-known spin Hamiltonians, namely, (1) the $XY$ model in a transverse 
field \cite{sup_xymodel,sup_xybooks}, (2) the $XY$ model in a longitudinal field, and (3) the $XXZ$ model in an external field \cite{sup_xxzmodel}.

\subsection{XY model in a transverse field}
\label{transversexy}

The two-qubit Hamiltonian describing the $XY$ model in a transverse field is given by \cite{sup_xymodel,sup_xybooks} 
\begin{eqnarray}
 H_{XY}&=&J\left[\frac{1+\gamma}{2}\sigma^{x}_{1}\sigma^{x}_{2}+\frac{1-\gamma}{2}\sigma^{y}_{1}\sigma^{y}_{2}\right]
 +h\sum_{i=1}^{2}\sigma_{i}^{z},
 \label{xy-ham}
\end{eqnarray}
where $J$ is proportional to the interaction strength, $\gamma$ is the anisotropy parameter, and $h$ is the strength of the external 
transverse magnetic field. 
Note that $H_{XY}$ can be written as a sum of an interacting and a local Hamiltonian as
\begin{eqnarray}
H_{XY}=H_{XY}^{int}+\frac{h}{J}H_{XY}^{l}, 
\end{eqnarray}
where 
\begin{eqnarray}
 H_{XY}^{int}=\left[\frac{1+\gamma}{2}\sigma^{x}_{1}\sigma^{x}_{2}+\frac{1-\gamma}{2}\sigma^{y}_{1}\sigma^{y}_{2}\right];
 H_{XY}^{l}=\sum_{i=1}^{2}\sigma_{i}^{z}. \nonumber \\
\end{eqnarray}
The average energy of an arbitrary two-qubit state, $\rho$, in the present 
case, has lower and upper bounds determined by the minimum and maximum eigenvalues of the Hamiltonian $H_{XY}$.  
We find that, for a given value of $h/J$, $-E^{\prime}\leq\mbox{tr}(H_{XY}\rho)\leq E^{\prime}$, where 
$E^{\prime}=\max\{1,\sqrt{4(h/J)^{2}+\gamma^{2}}\}$ for $0\leq\gamma<1$, and 
$E^{\prime}=\sqrt{4(h/J)^{2}+\gamma^{2}}$ for $\gamma\geq1$. Here and in the rest of the paper, the calculations of 
energy are always performed after a division of the Hamiltonian by the coupling constant, $J$, so that the ensuing energy 
expressions are dimensionless.

\begin{figure*}
 \includegraphics[scale=0.7]{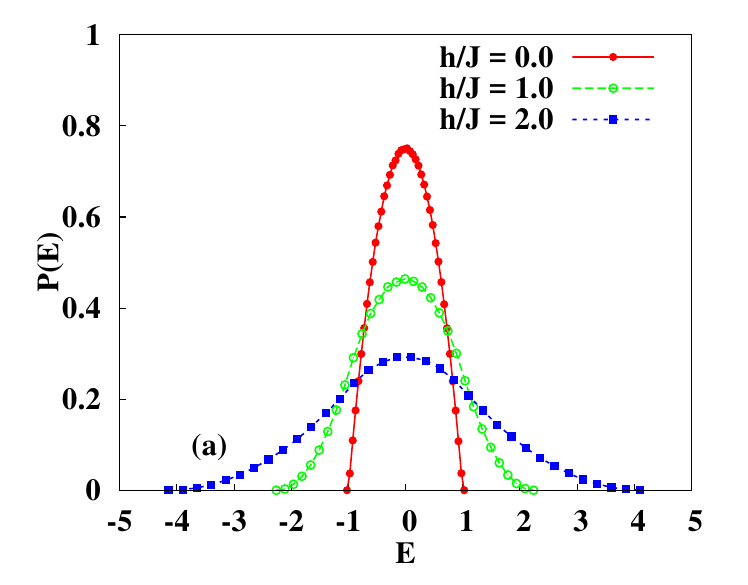}
 \includegraphics[scale=0.7]{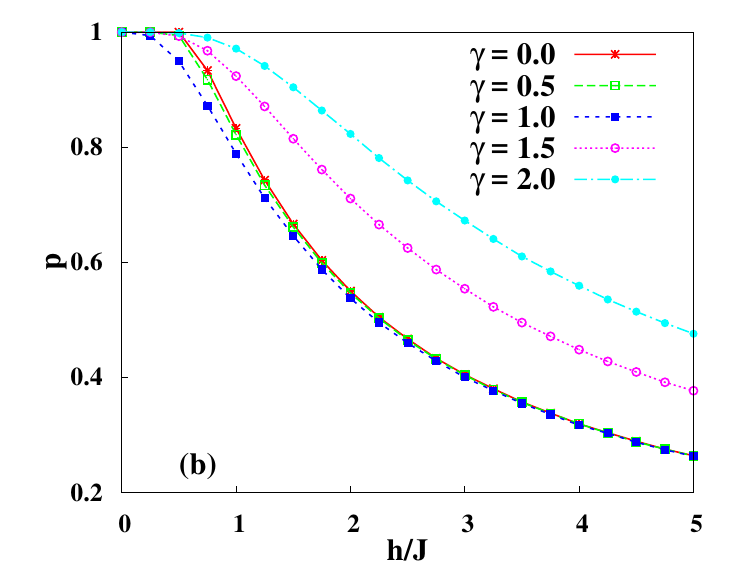}
 \caption{(Color online.) Special canonically distillable states with respect to the transverse-field $XY$ model. 
 (a) The probability distribution, $P(E)$, of average energy, $E$, 
 of Haar uniformly chosen random two-qubit pure states
 for different values of $h/J$ at $\gamma=1$. The distributions 
 becomes sharply peaked at $E=0$ for low values of $h/J$. 
 (b) The variation of the probability $p$ that a two-qubit pure state is SCD
 as a function of $h/J$ for different values of $\gamma$ (See Eq. (\ref{p:general})). In this and in all the following figures in 
 this paper, the calculation of $p$ does not use the assumption of independence. All quantities employed are dimensionless.}
 \label{xypure}
\end{figure*}

\begin{figure*}
 \includegraphics[scale=0.7]{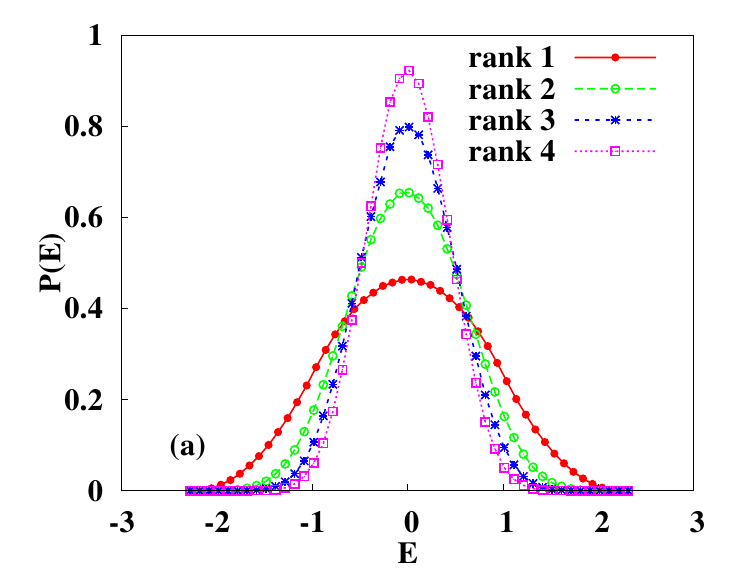}
 \includegraphics[scale=0.7]{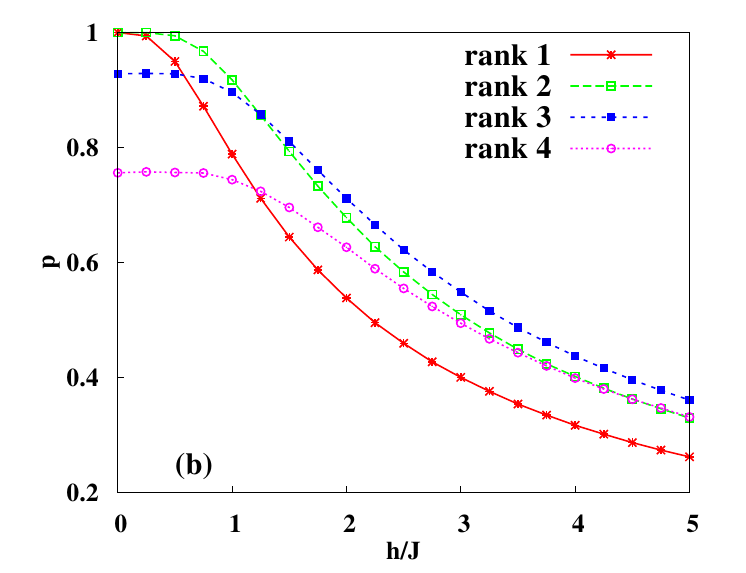}
 \caption{(Color online.) The effect of rank on special canonical distillability with respect to the transverse-field $XY$ model. 
 (a) The probability distribution, $P(E)$, of average energy, $E$, for Haar uniformly chosen  random mixed states of 
 different ranks with $h/J=1$, $\gamma=1$. The distribution 
 becomes sharply peaked at $E=0$ when the rank of the states increase. (b) The variation of the probability $p$ as a function of 
 $h/J$ at $\gamma=1$ for states of different ranks. Other considerations are the same as in Fig. \ref{xypure}.}
 \label{xymixed}
\end{figure*}

From weak canonical energy constraint (WCEC) and using Eq. (\ref{loc-zero}), we have $\langle\tilde{\psi}^{-}|H_{XY}|\tilde{\psi}^{-}\rangle$ $=$ 
$\langle\tilde{\psi}^{-}|H_{XY}^{int}|\tilde{\psi}^{-}\rangle$, which has lower and upper bounds given by the minimum and maximum 
eigenvalues of $H_{XY}^{int}$. In the present case, $-\epsilon\leq\langle\tilde{\psi}^{-}|H_{XY}|\tilde{\psi}^{-}\rangle\leq\epsilon$,
where $\epsilon=1$ for $0\leq\gamma<1$, and $\epsilon=\gamma$ for $\gamma\geq 1$.
The local unitary operator $U^{k}_{2}$, involved in obtaining \(|\tilde{\psi}^-\rangle\) 
for the qubit $k$ ($k=1,2$) can be parametrized as
\begin{eqnarray}
 U^{k}_{2}=\left( 
 \begin{array}{cc}
 \cos\theta^{k}e^{i\phi^{k}_{1}} & \sin\theta^{k}e^{i\phi^{k}_{2}} \\
 -\sin\theta^{k}e^{-i\phi^{k}_{2}} & \cos\theta^{k}e^{-i\phi^{k}_{1}} \\
 \end{array}
 \right).
 \label{unitary}
\end{eqnarray}
Since $\langle\tilde{\psi}^{-}|H_{XY}^{int}|\tilde{\psi}^{-}\rangle$ is a continuous function of  
$\{\theta^{k},\phi^{k}_{1},\phi^{k}_{2}\}$, the parameters of unitary operators with $k=1,2,$ one can always find at least one
unitary operator $\mathcal{U}_{2}$ for every value of $\langle\tilde{\psi}^{-}|H_{XY}^{int}|\tilde{\psi}^{-}\rangle$ in the 
allowed range $[-\epsilon,\epsilon]$.  
For example, for $\gamma>1$, choosing $\phi^{1}_{1}=\phi^{2}_{1}=\pi/2$, and $\phi^{1}_{2}=\phi^{2}_{2}=0$ results in 
$\langle\tilde{\psi}^{-}|H_{XY}|\tilde{\psi}^{-}\rangle=\frac{1}{2}(-1-\gamma+(-1+\gamma)\cos2(\theta_{1}-\theta_{2}))$, 
which can
exhaust the range $[-\gamma,-1]$ when $\theta_{1}$ and $\theta_{2}$ are varied. Besides, a choice of
$\phi^{k}_{j}=0$, $j,k=1,2$ gives 
$\langle\tilde{\psi}^{-}|H_{XY}|\tilde{\psi}^{-}\rangle=\frac{1}{2}(-1+\gamma-(1+\gamma)\cos2(\theta_{1}-\theta_{2}))$ by which 
the range $[-1,\gamma]$ can be exhausted in a similar fashion.
We therefore find that the inclusion of the interaction terms has a drastic effect on canonical distillability. While 
almost no states are SCD for $J=0$ (Proposition \ref{pr:localham}), a significant fraction of them are so for $J\neq 0$.

We have performed extensive numerical simulations to check the assumption of independence, and have found it to hold, 
within numerical accuracy, for the transverse $XY$ Hamiltonian. Therefore, via Proposition \ref{pr:int_term},  
the probability of an arbitrary two-qubit entangled state to be SCD with respect to $H_{XY}$ 
is given by $p=\eta\int_{-\epsilon}^{\epsilon}P(E)dE$.

\begin{figure*} 
 \includegraphics[width=0.55\textwidth]{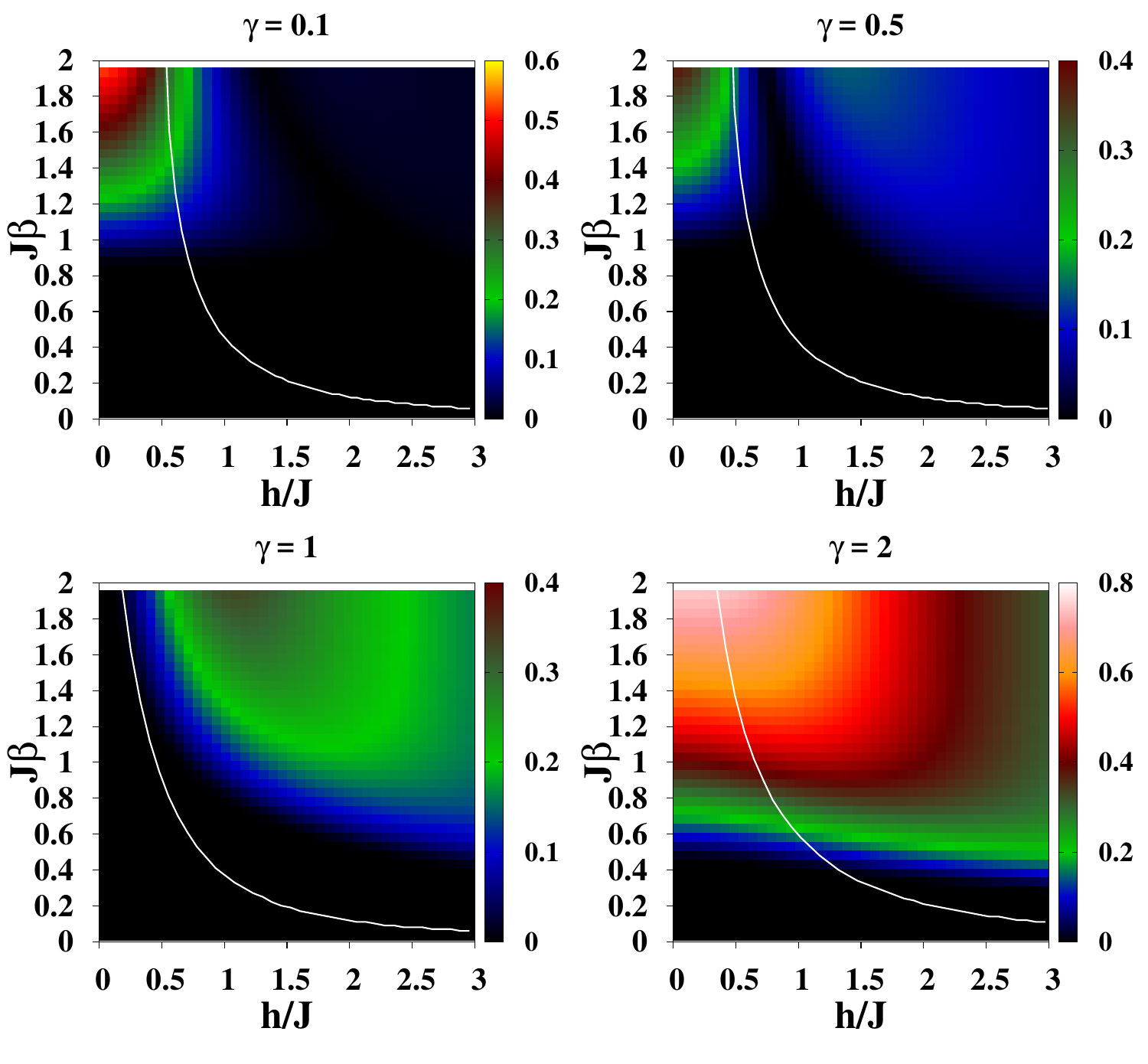}
 \caption{(Color online.) Canonical distillation phase diagrams for different values of $\gamma$ in the case of $XY$ model in a 
 transverse field. The white line depicts the boundary between thermal states
 that do and do not satisfy the WCEC on the $(h/J,J\beta)$ plane. 
 Below the boundary, the constraint is satisfied while above the boundary, it is not. The different shades in the figures 
 represent different values of entanglement as measured by concurrence as a function of $h/J$ and $J\beta$.
 All quantities are dimensionless, except the concurrence, which is in ebits.}
 \label{xythermal} 
\end{figure*}

 \begin{figure*} 
 \includegraphics[width=1\textwidth]{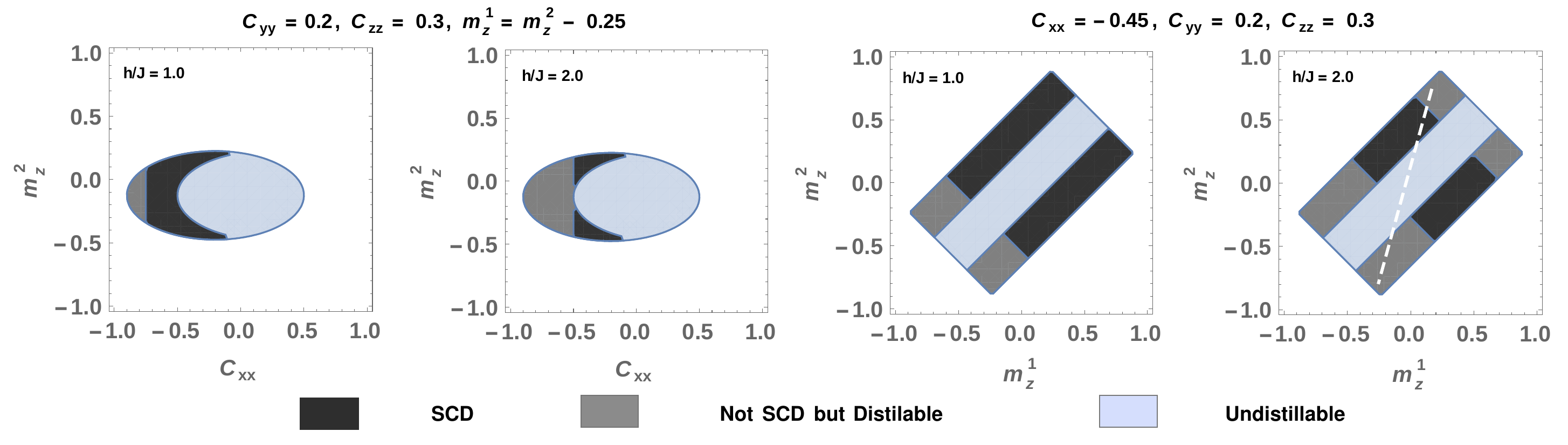}
 \caption{Canonical distillability of two-qubit states with local magnetizations. 
 The projections of the SCD region on different two-dimensional cross-sections of the 
 parameter space of the state \(\rho_{m}\), where the system is governed by the two-qubit \(XY\) Hamiltonian. 
 The shaded regions depict the projections of the 
 volume of physical states on the different cross-sections. 
 The first two figures consider projections onto the $(c_{xx},m_{z}^{2})$ plane, while the next two figures consider that onto the 
 $(m_{z}^{1},m_{z}^{2})$ plane. On the other hand, the 
 first and third figures are for $h/J=1.0$, while the rest are for $h/J=2.0$. 
 The black regions depict the states that are SCD, while the dark gray regions 
 depict the states that are not SCD but distillable in the usual, non-canonical, sense. 
 The light gray regions indicate undistillable 
 quantum states. Note that the region of SCD states diminishes with the increase in the value of the external field. This is 
 consistent with Proposition \ref{pr:localham}  and the fact that the probability, $p$, of a state being SCD decreases with increasing 
 external field (see Fig. \ref{xymixed}). 
 Also, an SCD state can be obtained by mixing two quantum state that are distillable, but not SCD, as  
 indicated by the dotted line in the rightmost figure, implying the non-convexity of the SCD states.
 }
 \label{bellvolume}
 \end{figure*}

Irrespective of the value of $h$, in the limit $\gamma\rightarrow\infty$, 
the probability of an arbitrary entangled state of a two-qubit system described by the Hamiltonian $H_{XY}$ to be SCD 
is unity. This can be understood by noting that $|E^{\prime}|\rightarrow|\epsilon|$ as $\gamma\rightarrow\infty$. 
Also, note that for $0\leq\gamma<1$, $E^{\prime}\rightarrow1$ when $h\rightarrow0$. Since $\epsilon=1$ for $0\leq\gamma<1$, 
we conclude that $p\rightarrow 1$ in the limit $h\rightarrow 0$ for $0\leq\gamma<1$. 
On the other hand, for $\gamma\geq1$, $E^{\prime}\rightarrow\gamma$ when $h\rightarrow0$. 
And $\epsilon=\gamma$ for $\gamma\geq1$.  
This leads to the following proposition.

\begin{proposition}
 In a two-qubit system described by the transverse \(XY\) Hamiltonian, 
all entangled states are SCD, provided either \(h \rightarrow 0\) or \(\gamma \rightarrow \infty\).
\label{pr:twoqubitxy}
\end{proposition}

The probability distributions, $P(E)$, of average energy of the state over the space of all two-qubit pure states 
for different values of $h/J$ are depicted  
in Fig. \ref{xypure}(a). The variation of the probability $p$ with the field-strength $h$ in the case of pure states 
of the two-qubit $XY$ model is depicted in Fig. \ref{xypure}(b), for different values of $\gamma$. Note that the value of $p$ 
decreases as $h/J$ increases, and asymptotically vanishes as $h\rightarrow\infty$. This can be understood 
as a result of Proposition \ref{pr:localham}.  
For a fixed $h/J$, $p$ is a non-monotonic function of $\gamma$ with a change in character at $\gamma=1$, the Ising point.  
For $\gamma<1$, $p$ decreases with increasing $\gamma$ for a fixed value of $h$ while for 
$\gamma>1$, $p$ increases as $\gamma$ increases, as clearly seen from Fig. \ref{xypure}(b).

In the case of two-qubit mixed states, we find that the DF, $\eta$, is a function of the rank, $r$, of the state, 
with $\eta(r=2)=1$, $\eta(r=3)=0.928$, and $\eta(r=4)=0.756$. 
These estimates are obtained via numerical simulation, for which we have generated $10^{8}$ states (for each rank) Haar 
uniformly over the space of quantum states of the corresponding rank. 
The probability distributions, $P(E)$, of average energy of the 
two-qubit entangled mixed states with different ranks are shown in Fig. \ref{xymixed}(a) for $h/J=1$, $\gamma=1$. As in the case of 
pure states, the probability distributions are bell-shaped, and becomes sharply peaked around zero for states with higher ranks. 
Fig. \ref{xymixed}(b) depicts the variation of $p$ against $h$ for mixed 
states of rank $2$, $3$, and $4$ with $\gamma=1$. Similar to the case of pure states, the probability $p\rightarrow0$ as 
$h\rightarrow\infty$.  

Let us now discuss the canonical distillability of some special types of mixed states.

\noindent\textbf{\emph{Thermal states.--}} We intend to find out whether canonical distillability of a two-qubit mixed entangled state 
depends on temperature. To pursue this question, we construct the thermal state of the two-qubit system as 
$\rho_{th}=\exp(-\beta H_{XY})/Z$, where $Z=\mbox{tr}\{\exp(-\beta H_{XY})\}$ is the partition function of the system
and $\beta=1/k_{B}T$, $k_{B}$ and $T$ being the Boltzmann constant and absolute temperature respectively. 
Fig. \ref{xythermal} depicts the canonical distillation
phase diagram on the $(h/J,J\beta)$ plane for different values of $\gamma$ in case of the $XY$ model in a transverse field.
For every value of $h$, there is a critical value of $T$ above which the state satisfies 
the WCEC (Eq. (\ref{cdeq})),
whereas below the critical value, it does not.  We call this value as the \emph{SCD temperature}. Note that in Fig. \ref{xythermal}, the 
vertical axis is proportional to $\beta$, i.e., proportional to inverse of $T$. 
For the zero temperature cases, we find that increase of entanglement tends to facilitate canonical distillability. However, after thermal mixing, 
there is a trade-off between temperature and entanglement, which aids in canonical distillability at higher temperatures where 
entanglement is typically low. 
The entanglement
of the state, as measured by concurrence \cite{sup_eform}, is mapped onto the $(h/J,J\beta)$ plane and is represented by different shades in 
Fig. \ref{xythermal}. Note that for low values of $h/J$, the thermal states have low SCD temperatures,
whereas the trend is opposite for higher values of $h/J$.  
The figure clearly shows that there exist thermal states that are not entangled, but satisfies Eq. (\ref{cdeq}). Similarly, 
thermal entangled states exist which do not satisfy Eq. (\ref{cdeq}) and hence can not be SCD.

\noindent\textbf{\emph{Bell-diagonal states.--}} Next, we explore the canonical distillability of Bell-diagonal (BD) states given by 
\begin{eqnarray}
\rho_{BD}=\frac{1}{4}\{I_{1}\otimes I_{2}+\sum_{\alpha}c_{\alpha\alpha}\sigma_{1}^{\alpha}\otimes\sigma_{2}^{\alpha}\},
\end{eqnarray} 
with $\alpha=x,y,z$,
where $c_{\alpha\alpha}=\mbox{tr}(\sigma_{1}^{\alpha}\otimes\sigma_{2}^{\alpha}\rho_{BD})$ are the diagonal elements of the 
correlation matrix 
($|c_{\alpha\alpha}|\leq1$), and 
where $I_{1}$ and $I_{2}$ are the identity matrices in the Hilbert spaces of the qubits $1$ and $2$, respectively. 
The positivity of the BD state dictates that the correlators, $\{c_{\alpha\alpha}\}$, are 
constrained to vary within a strict subset of the hypercube.
The average energy of the state is given by $E^{XY}_{BD}=((1+\gamma)c_{xx}+(1-\gamma)c_{yy})/2$.                                 
Hence, $E^{XY}_{BD}$ $\in$ $[-1,1]$ for $0\leq\gamma<1$, and $E^{XY}_{BD}$ $\in$ $[-\gamma,\gamma]$ when 
$\gamma\geq 1$. Since the range of the left-hand side of Eq. (\ref{cdeq}) coincides with that of the right-hand side,
the BD states are always SCD for the transverse $XY$ model. 

\noindent\textbf{\emph{Mixed states with fixed magnetizations.--}} Introduction of magnetization in the $z$-direction to the state
$\rho_{BD}$ makes the two-qubit mixed state to be of the form 
\begin{eqnarray}
\rho_{m}=\frac{1}{4}\{I_{1}\otimes I_{2}+\sum_{\alpha}c_{\alpha\alpha}\sigma_{1}^{\alpha}\otimes\sigma_{2}^{\alpha}
+m_{z}^{1}\sigma_{1}^{z}\otimes I_{2}+m_{z}^{2}I_{1}\otimes\sigma_{2}^{z}\}.
\end{eqnarray} 
Here, $m_{z}^{1}$ and $m_{z}^{2}$ represent
the magnetizations of the qubits $1$ and $2$, respectively, with $m_{z}^{i}=\mbox{tr}(\sigma^{z}_{i}\rho_{i})$, $i=1,2$, 
$\rho_{i}$ being the local density matrix of the qubit $i$. Similar to the correlators $c_{\alpha\alpha}$, $-1\leq m_{z}^{i}\leq 1$.
The average energy of this state is given by $E^{XY}_{m}=\frac{h}{J}(m_{z}^{1}+m_{z}^{2})+((1+\gamma)c_{xx}+(1-\gamma)c_{yy})/2$. The 
canonical distillability of the state depends explicitly on the values of the correlators and the magnetizations. 
Fig. \ref{bellvolume} depicts the projections of the volume in the $(c_{xx},m_{z}^{2})$ and $(m_{z}^{1},m_{z}^{2})$ planes, 
in which all the states of the form $\rho_{m}$ are SCD (red regions) for fixed values of the other parameters 
($c_{yy}=0.2$, $c_{zz}=0.3$). With increasing $h/J$, the region shrinks, thereby indicating a decrease in the probability that a 
mixed state of the form $\rho_{m}$ is distillable and SCD. This is consistent with our earlier findings regarding  the dependence 
of special canonical distillability of mixed entangled states on $h/J$.

\begin{figure*} 
 \includegraphics[width=0.6\textwidth]{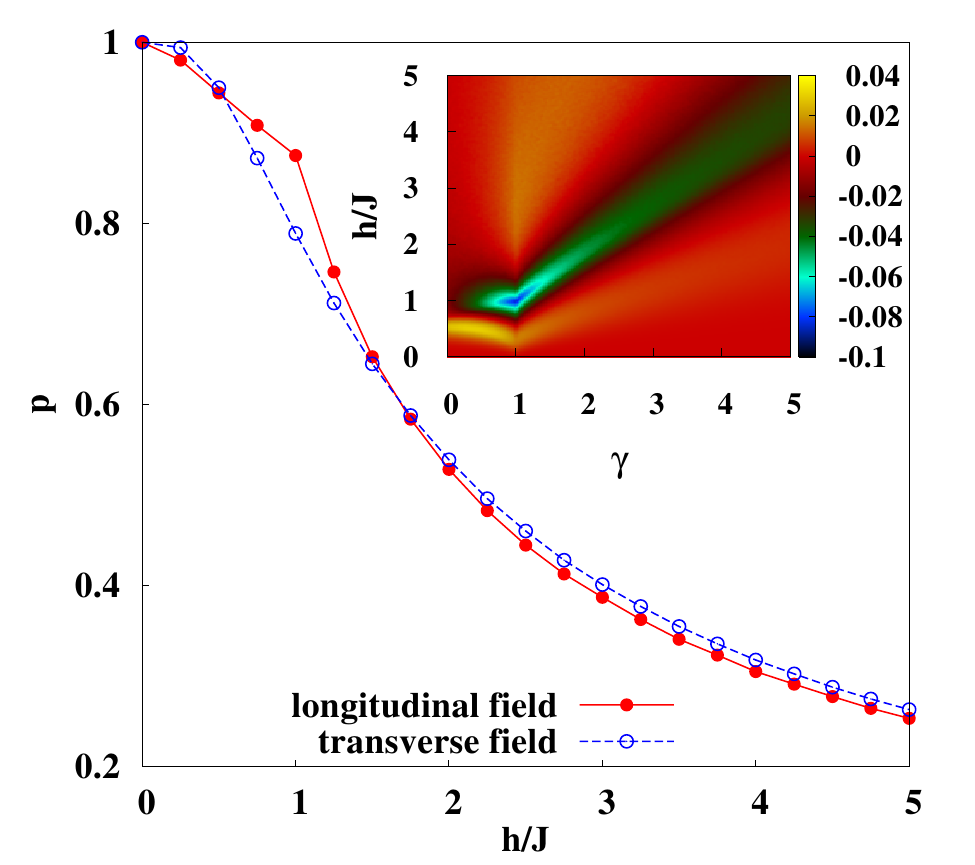}
 \caption{(Color online.) The variation of the probability $p$, for pure states, as functions of $h/J$ in the 
 transverse- and longitudinal-field Ising models. The rate of decay of $p$ changes abruptly at $h/J=1$ in the case of the 
 longitudinal field. The inset exhibits the difference between the probabilities for the two models. See text 
 (Sec. \ref{longitudexy})
 for details. All quantities plotted are dimensionless.
 }
 \label{longxy} 
\end{figure*}

\subsection{XY model in a longitudinal field}
\label{longitudexy}

To investigate whether a change in the direction of the external field alters the probability of a state to be SCD,  
we consider the two-qubit $XY$ model in a longitudinal field, described by the Hamiltonian
\begin{eqnarray}
 H_{XY(l)}&=&J\left[\frac{1+\gamma}{2}\sigma^{x}_{1}\sigma^{x}_{2}+\frac{1-\gamma}{2}\sigma^{y}_{1}\sigma^{y}_{2}\right]
 +h\sum_{i=1}^{2}\sigma_{i}^{x},\nonumber \\
 \label{xyl-ham}
\end{eqnarray}
where $h$ is the strength of the longitudinal field. Note that we are using the same symbol $h$ for the longitudinal field that
we had used in the preceding case for the transverse field.

The probability $p$, in the case of the longitudinal-field Ising model $(\gamma=1)$, is plotted against $h/J$ in 
Fig. \ref{longxy}. Note that the rate of decay
of $p$ with $h$ changes abruptly at $h/J=1$. This can be understood by noting that the average energy of an entangled state
of a two-qubit system described by the Hamiltonian $H_{XY(l)}$ for $\gamma=1$ is bounded below and above by the minimum and maximum 
eigenvalues of the Ising Hamiltonian, respectively. They are given by $E_{1}^{XY(l)}=-1$, $E_{2}^{XY(l)}=\frac{2h}{J}+1$ for 
$\frac{h}{J}<1$, and $E_{1}^{XY(l)}=-\frac{2h}{J}+1$, $E_{2}^{XY(l)}=\frac{2h}{J}+1$ for $\frac{h}{J}\geq1$. Due to the  
energy level crossing  at $\frac{h}{J}=1$, the allowed
range of values for the average energy changes whereas the range of $\langle\tilde{\psi}|H_{XY}^{l}|\tilde{\psi}\rangle$ 
remains fixed at $[-1,1]$ irrespective of the values of $\frac{h}{J}$, thereby changing variation of $p$ abruptly. 
In Fig. \ref{longxy}, we 
compare the result of longitudinal-field Ising model with that obtained from the transverse-field Ising model where the same 
field strength is applied in the transverse direction. We observe 
that over a certain interval of the field values, the probability is greater in the case of the longitudinal field than that in the 
case of the transverse field, while the opposite is true in the complementary region. This
 clearly indicates that the probability depends on the direction of the applied field. 
For comparison between the longitudinal and transverse models, we introduce the quantity
$\Delta p=p(H_{XY})-p(H_{XY(l)})$, and plot it as a function of $h/J$ and $\gamma$ in the inset of Fig. \ref{longxy}. 
The existence of both positive and negative values of $\Delta p$ 
reveals the interesting fact that $p$ can be increased or decreased by changing only the direction of the field.

\subsection{XXZ model in an external field}
\label{xxzinfield}

The two-qubit $XXZ$ model in an external field is described by the Hamiltonian \cite{sup_xxzmodel}
\begin{eqnarray}
 H_{XXZ}=\frac{J}{2}(\sigma_{1}^{x}\sigma_{2}^{x}+\sigma_{1}^{y}\sigma_{2}^{y}+\Delta\sigma_{1}^{z}\sigma_{2}^{z})
  +h\sum_{i=1}^{2}\sigma_{i}^{z},
 \label{xxz-ham}
\end{eqnarray}
where $J$ is proportional to the interaction strength, $\Delta$ is the anisotropy in $z$ direction, 
and $h$ is the strength of the external field.
The average energy of any entangled state of a two-qubit system defined by the Hamiltonian $H_{XXZ}$ must be in 
the range $[E_{1}^{XXZ},E_{2}^{XXZ}]$, where $E_{1}^{XXZ}$ and $E_{2}^{XXZ}$ are the minimum and maximum eigenvalues, respectively, 
of $H_{XXZ}$, with $E_{1}^{XXZ}$, $E_{2}^{XXZ}$ $\in$ $\{-1-\Delta/2,1-\Delta/2,
-2(h/J)+\Delta/2,2(h/J)+\Delta/2\}$. The choice of $E_{1}$ and $E_{2}$ depends on the values of $\Delta$ and $h/J$. 

\begin{figure*}
 \includegraphics[width=0.6\textwidth]{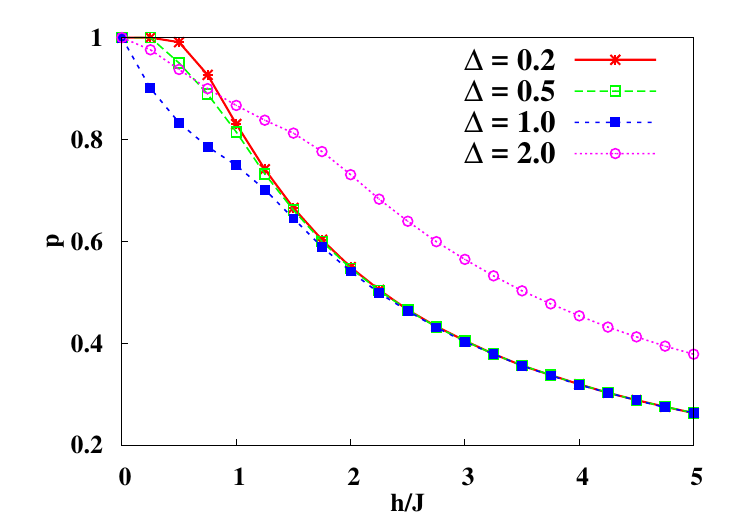}
 \caption{(Color online.) The variation of the probability $p$, for pure states,  
 as a function of $h/J$ for different values of $\Delta$ in the case of 
 two-qubit $XXZ$ model in an external field. All quantities employed are dimensionless.}
 \label{xxzpure}
\end{figure*}

The right hand side of the WCEC (Eq. (\ref{cdeq})) lies in the range $[\epsilon_{1},\epsilon_{2}]$, 
where $\epsilon_{1}$ and $\epsilon_{2}$ are the minimum and maximum eigenvalues, respectively, of the interacting part of the 
$XXZ$ Hamiltonian given by $H_{XXZ}^{int}=\frac{J}{2}(\sigma_{1}^{x}\sigma_{2}^{x}+\sigma_{1}^{y}\sigma_{2}^{y}
+\Delta\sigma_{1}^{z}\sigma_{2}^{z})$. For all values of $\Delta$, $\epsilon_{1}=-(1+\Delta/2)$ whereas 
$\epsilon_{2}=1-\Delta/2$ for $0\leq\Delta<1$, and $\epsilon_{2}=\Delta/2$ for $\Delta\geq1$. 
The probability of a two-qubit 
state being SCD can be obtained following Proposition \ref{pr:int_term}. Note that with 
unitary operators of the form (\ref{unitary}), one can obtain all the  values of
$\langle\tilde{\psi}^{-}|H_{XXZ}|\tilde{\psi}^{-}\rangle$ in the allowed range $[-\epsilon_{1}^{XXZ},\epsilon_{2}^{XXZ}]$ since 
$\langle\tilde{\psi}^{-}|H_{XXZ}^{int}|\tilde{\psi}^{-}\rangle$ is a continuous function of the parameters 
$\{\theta^{j},\phi^{j}_{1},\phi^{j}_{2}\}$, $j=1,2$. Fig. \ref{xxzpure} represents the decay of the 
probability $p$ as a function of $h/J$ for different values of $\Delta$ in the case of pure states. Note that the dependence 
of $p$ on $\Delta$, similar to that of $p$ on $\gamma$ in the case of the longitudinal-field $XY$ model, is non-monotonic for a 
fixed value 
of $h/J$. When $\Delta<1$, $p$ decreases with increasing $\Delta$ for a fixed $h/J$ whereas for $\Delta>1$, the opposite trend is 
observed. Also, the decay rate of $p$  with $h/J$ changes abruptly for $\Delta\geq1$ due to the ground state energy level crossing
that changes the limits of the distribution $P(E)$, similar to the case of the $XY$ model in a longitudinal field. In the case of 
two-qubit mixed states, one can find features similar to that in the case of transverse-field $XY$ 
Hamiltonian (as depicted in Fig. \ref{xymixed}) using the same methodology. 

\begin{figure*} 
 \includegraphics[width=0.6\textwidth]{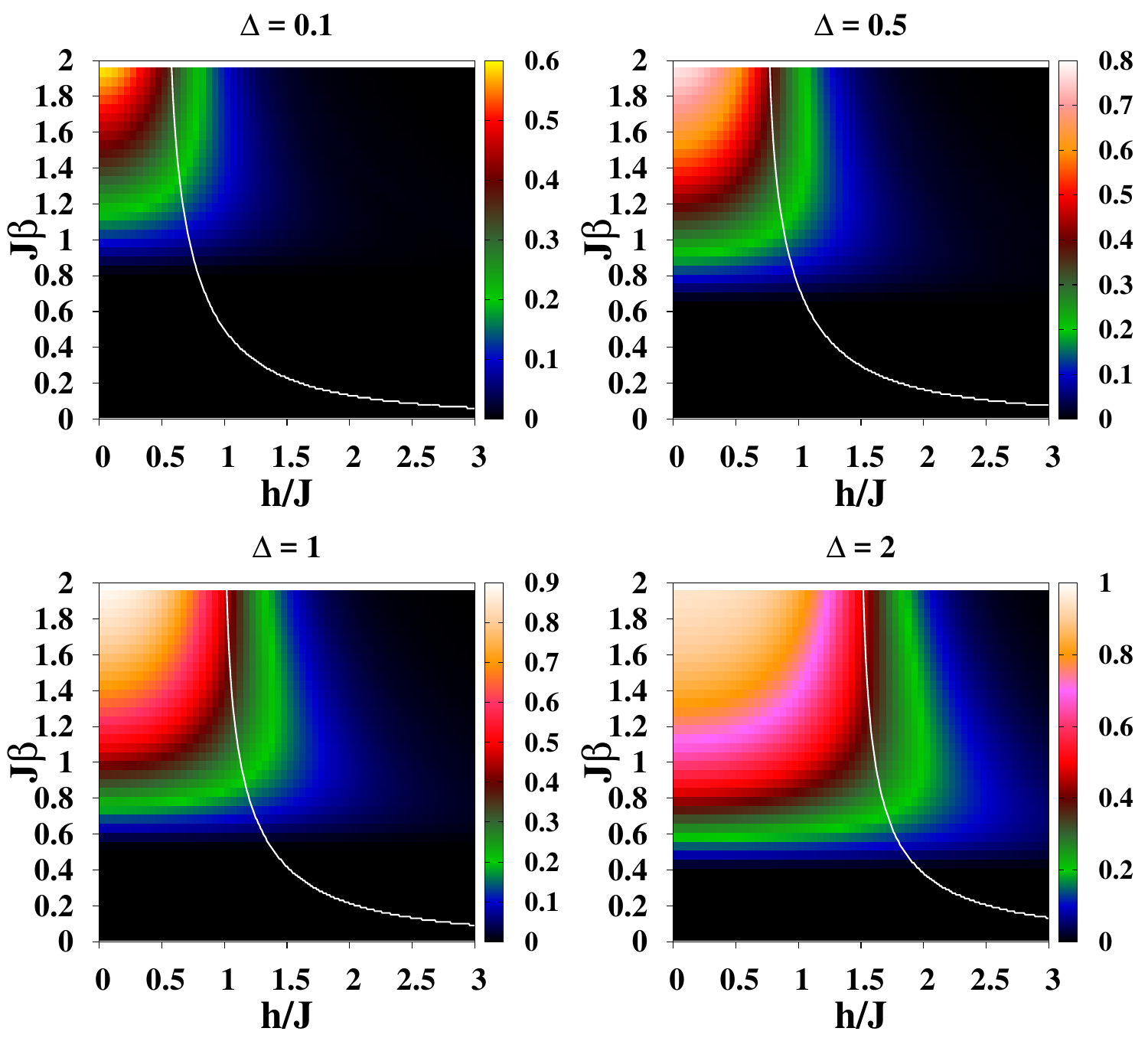}
 \caption{(Color online.) Canonical distillability phase diagrams for different values of $\Delta$ in the case of $XXZ$ model 
 in an external field. All other considerations are the same as in Fig. \ref{xythermal}.}
 \label{xxzthermal} 
\end{figure*}

We conclude the discussion on the $XXZ$ model by examining the canonical distillability of the thermal states constructed from 
the eigen-spectrum of the model. We find that similar to the transverse-field $XY$ model, a critical temperature exists 
for all values of $h/J$, for 
a fixed $\Delta$, above which the thermal state satisfies Eq. (\ref{cdeq}), but below which it does not. 
(See 
Fig. \ref{xxzthermal}). The zero-entanglement regions above the white line indicates that there exist thermal states that  
satisfy Eq. (\ref{cdeq}), but are non-distillable.

\section{Two-Qutrit systems}
\label{twoqutrit}

\begin{figure*}
 \includegraphics[width=0.5\textwidth]{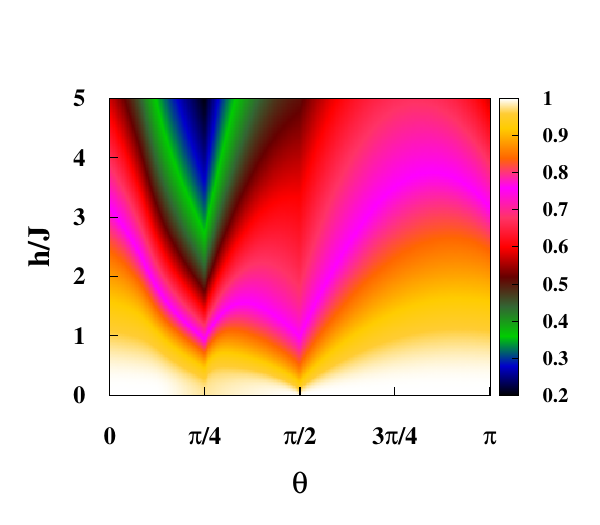}
 \caption{(Color online.) The probability, $p$, for pure states, as a function of the interaction parameter $\theta$ and 
 the field-strength $h/J$ for the two-qutrit model described by the Hamiltonian $H_{3,3}$. 
 A general two-qutrit state is less possible to be SCD with the increase of $h/J$ for a fixed value of $\theta$.
 All quantities used are dimensionless.}
 \label{qutrit}
\end{figure*}

We now conclude an example of a two-qutrit system 
defined by a bilinear-biquadratic Hamiltonian \cite{sup_bbh} in the presence of a field term. The Hamiltonian is given by  
\begin{eqnarray}
 H_{3,3}=J\left[\cos\theta\vec{S}_{1}.\vec{S}_{2}+\sin\theta\left(\vec{S}_{1}.\vec{S}_{2}\right)^{2}\right]
 +h\sum_{i=1}^{2}S_{i}^{z},
\end{eqnarray}
where $\vec{S}_{i}=\left\{S_{i}^{x},S_{i}^{y},S_{i}^{z}\right\}$, $i=1,2$, are the spin operators on the qutrit $i$, with 
\begin{eqnarray}
 S_{i}^{x}&=&\frac{1}{\sqrt{2}}
 \left( 
 \begin{array}{ccc}
  0 & 1 & 0 \\
  1 & 0 & 1 \\
  0 & 1 & 0
 \end{array}
 \right),
  S_{i}^{y}=\frac{1}{\sqrt{2}i}
 \left( 
 \begin{array}{ccc}
  0 & 1 & 0 \\
  -1 & 0 & 1 \\
  0 & -1 & 0
 \end{array}
 \right), \nonumber \\
  S_{i}^{z}&=&\mbox{diag}\{1,0,-1\}.
\end{eqnarray}
Here, $\cos\theta$ and $\sin\theta$ are the relative strengths of the bilinear and biquadratic interactions, respectively. 
The probability that a randomly chosen two-qutrit state $\rho$ is SCD is determined by an equation similar to Eq. (\ref{cdeq})
where $|\psi^{-}\rangle$ is replaced by $|\Phi\rangle$ (Eq. (\ref{qudit_state})) with $d=3$. The left hand side 
of the equation represents
the average energy of the two-qutrit state which is bounded below and above by the minimum and the maximum eigenenergy 
of the Hamiltonian, given by $E_{1}^{3,3}$ and $E_{2}^{3,3}$, respectively. Here, $E_{1}^{3,3}=\min\{\mathcal{S}\}$ and 
$E_{2}^{3,3}=\max\{\mathcal{S}\}$ where   
$\mathcal{S}=\{\cos\theta+\sin\theta,-\cos\theta+\sin\theta,-2\cos\theta+4\sin\theta,-h/J-\cos\theta+\sin\theta,h/J-\cos\theta+\sin\theta,
-2h/J+\cos\theta+\sin\theta,-h/J+\cos\theta+\sin\theta,h/J+\cos\theta+\sin\theta,2h/J+\cos\theta+\sin\theta\}$ is the 
set of eigenvalues of
the Hamiltonian, the maximum and the minimum being determined by the values of $h/J$ and $\theta$. 
The probability of a randomly chosen two-qutrit pure state to be SCD is plotted on the $(\theta,h/J)$ plane in Fig. \ref{qutrit}. 
The probability decreases as $h/J$ increases for a fixed value of $\theta$, as consistent with 
Proposition \ref{pr:localham}. Note that 
at $h=0$, the probability $p$, unlike in the case of the two-qubit pure states, is not always unity but depends on $\theta$ and 
therefore on the strengths of the bilinear and biquadratic interactions. 

\section{Multipartite systems}
\label{cd-multiparty_supp}

Canonical distillability of a multipartite system constituted of $N$ parties and described by a Hamiltonian $H$ 
can be investigated using the same methodology as that used in the case of bipartite systems. 
For the purpose of demonstration, we restrict ourselves to three-qubit states 
belonging to the well-known  GHZ and W classes \cite{sup_ghzstate,sup_zhgstate,sup_dvc}, which are mutually disjoint sets that collectively 
exhaust the entire set of three-qubit pure states. Starting with three-qubit states chosen from the GHZ class as resource states, 
we consider the distillation of the multiparty entangled three-qubit GHZ state, 
$|\psi\rangle_{GHZ}=(|000\rangle+|111\rangle)/\sqrt{2}$, using the WCEC. Similarly, while considering the W 
class of states as the resource states, the target state is the three-qubit W state, 
$|\psi\rangle_{W}=(|001\rangle+|010\rangle+|100\rangle)/\sqrt{3}$. 
Let the multiqubit system be described by the $XY$ Hamiltonian in an external transverse field, given by 
\begin{eqnarray}
 H_{XY}^{N}&=&J\sum_{i=1}^{N}\left[\frac{1+\gamma}{2}\sigma_{i}^{x}\sigma_{i+1}^{x}
 +\frac{1-\gamma}{2}\sigma_{i}^{y}\sigma_{i+1}^{y}\right]
 +h\sum_{i=1}^{N}\sigma_{i}^{z}.\nonumber \\
 \label{xy_ham_n}
\end{eqnarray}

\subsection{GHZ class}
\label{ghzclass}

A general three-qubit state belonging to GHZ class is given by \cite{sup_ghzstate,sup_dvc}
\begin{eqnarray}
 |GHZ_{C}\rangle=\frac{1}{M}(a_{1}|0\rangle_{1}|0\rangle_{2}|0\rangle_{3}+
 a_{2}|\phi_{1}\rangle|\phi_{2}\rangle|\phi_{3}\rangle),\nonumber \\
\end{eqnarray}
where $M$ is a normalization constant, and $|\phi_{i}\rangle=b_{i}|0\rangle_{i}+c_{i}|1\rangle_{i}$, $i=1,2,3$, 
with $|b_{i}|^{2}+|c_{i}|^{2}=1$. Let us consider the case of $\gamma=1$ (transverse Ising model). 
One side of the WCEC in this case consists of the 
average energy of the three-qubit states in the 
GHZ class and has the minimum and maximum allowed values, $E_{1}^{N=3}$ and $E_{2}^{N=3}$, as the minimum and maximum eigenvalues, 
respectively, of $H_{XY}^{N=3}$. 
Choosing the standard state on the other side of the WCEC as $|\psi\rangle_{GHZ}$, 
we find that the part $h\sum_{i=1}^{3}\sigma_{i}^{z}$ of the Hamiltonian $H_{XY}^{N=3}$ does not contribute and the bounds on this 
side are decided by the minimum and maximum eigenvalues of $J\sum_{i=1}^{3}\left[\frac{1+\gamma}{2}\sigma_{i}^{x}\sigma_{i+1}^{x}
 +\frac{1-\gamma}{2}\sigma_{i}^{y}\sigma_{i+1}^{y}\right]$:   
\begin{eqnarray} 
-1\leq_{GHZ}\langle\tilde{\psi}|H_{XY}^{N=3}|\tilde{\psi}\rangle_{GHZ}\leq3, 
\end{eqnarray}
where $|\tilde{\psi}\rangle_{GHZ}$ is the state $|\psi\rangle_{GHZ}$ up to local unitary operators. 
With the choice of unitary operators as 
\begin{eqnarray}
 U_{1}=\left( 
 \begin{array}{cc}
 \cos\theta & \sin\theta \\
 -\sin\theta & \cos\theta \\
 \end{array}
 \right),
 U_{2}=U_{3}=\frac{1}{\sqrt{2}}\left( 
 \begin{array}{cc}
 1 & 1 \\
 -1 & 1 \\
 \end{array}
 \right),\nonumber \\
 \label{unitary3}
\end{eqnarray}
one obtains $_{GHZ}\langle\tilde{\psi}|H_{XY}^{N=3}|\tilde{\psi}\rangle_{GHZ}=1+2\sin\theta$, a continuous function, 
using which one can exhaust all possible real values in the allowed range $[-1,3]$ by varying $\theta$.  Therefore, the 
probability that a three-qubit state of the GHZ class is SCD to the three-qubit GHZ state can be obtained by 
determining the normalized probability 
distribution $P(E)$ for the average energy $E$ of the GHZ states and integrating it within proper limits: 
\begin{eqnarray}   
p=\int_{-1}^{3}P(E)dE. 
\end{eqnarray}
The DF is unity by virtue of Ref. \cite{sup_dvc}. 
The probability distribution 
$P(E)$ in the case of the three-qubit transverse Ising model and the GHZ class of states is shown in Fig. \ref{three}(a).
Fig. \ref{three}(b) shows the variation of the probability $p$ as a decreasing function of $\frac{h}{J}$.  

\subsection{W Class}
\label{wclassstates}

\begin{figure*} 
 \includegraphics[scale=0.35]{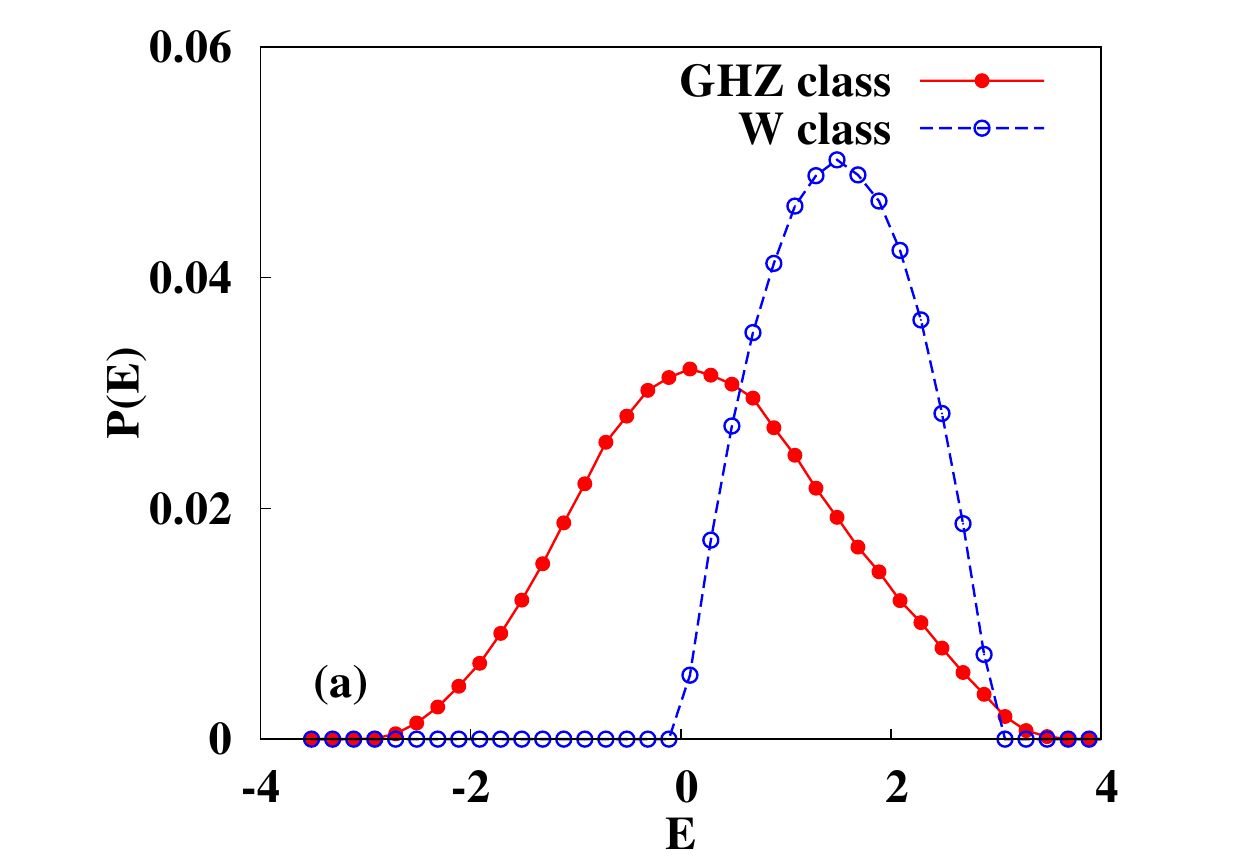}
 \includegraphics[scale=0.35]{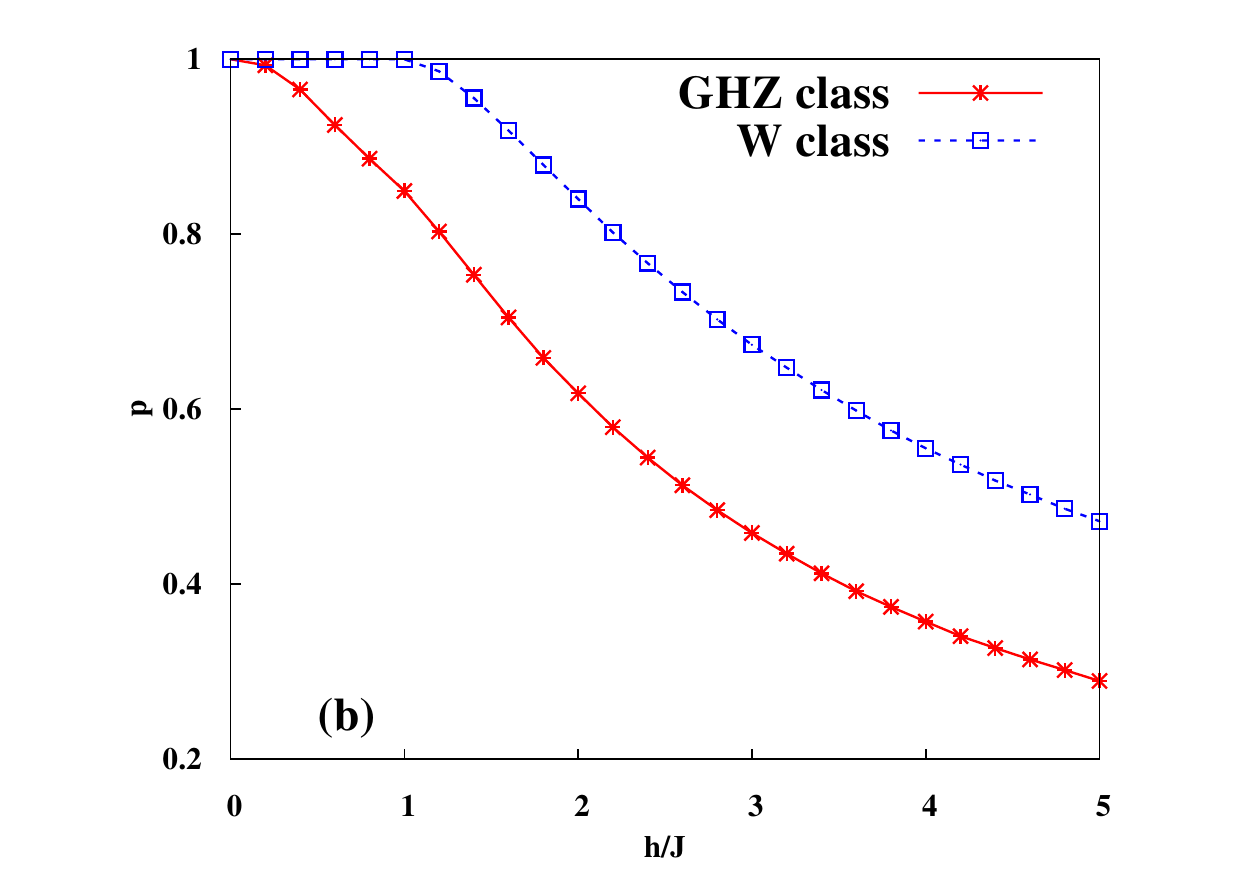}
 \caption{(Color online.) Multiparty special canonically distillable states. 
 (a) The probability distributions, $P(E)$, of the average energy, $E$, in the case of 
 three-qubit states belonging to the GHZ and W classes. The Hamiltonian parameters are set to $\gamma=1$ and $h/J=1$. 
 (b) The variation of the probability $p$ as a function of $h/J$ in the GHZ and W classes. All quantities employed are 
 dimensionless.}
 \label{three} 
\end{figure*}

A three-qubit state belonging to the W class is given by \cite{sup_zhgstate,sup_dvc}
\begin{eqnarray}
 |W_{C}\rangle=a|001\rangle+b|010\rangle+c|100\rangle+d|000\rangle,
\label{wclass} 
\end{eqnarray}
with $|a|^{2}+|b|^{2}+|c|^{2}+|d|^{2}=1$. We intend to canonically distill three-qubit W states, 
$|\psi\rangle_{W}$, where the system is described by the Hamiltonian $H_{XY}^{N=3}$. Similar to the GHZ class, we consider $\gamma=1$.
Both sides of the WCEC in this case are bounded by 
$E_{1}^{N=3}$ and $E_{2}^{N=3}$. 
However, we find that the effective range  is a strict subset of $[E_{1}^{N=3},E_{2}^{N=3}]$. Denoting the ends of that subset 
by $\epsilon_{1}^{W}$ and $\epsilon_{2}^{W}$, 
the probability $p$, in the present case, is given by 
$p=\int_{\epsilon_{1}^{W}}^{\epsilon_{2}^{W}}P(E)dE$, where the probability distribution $P(E)$ is exhibited in Fig. \ref{three}(a). 
The DF is again unity by virtue of Ref. \cite{dvc}. 
The variation of $p$ as a function of $h$ is shown in Fig. \ref{three}(b). 
Note that, in contrast to the case of GHZ class, up to $h/J=1$, 
we have $p=1$. 
When $h/J>1$, $p$ decreases with increasing $h/J$. 
This is therefore an occasion where it is advantageous to create W-class states than GHZ-class states (cf. \cite{W-vs-GHZ}).

\end{document}